# Distributed Random Access Algorithm: Scheduling and Congesion Control

L. Jiang    D. Shah    J. Shin    J. Walrand[*]

UCB and MIT

**Abstract**

This paper provides proofs of the rate stability, Harris recurrence, and $\varepsilon$-optimality of CSMA algorithms where the backoff parameter of each node is based on its backlog. These algorithms require only local information and are easy to implement.

The setup is a network of wireless nodes with a fixed conflict graph that identifies pairs of nodes whose simultaneous transmissions conflict. The paper studies two algorithms. The first algorithm schedules transmissions to keep up with given arrival rates of packets. The second algorithm controls the arrivals in addition to the scheduling and attempts to maximize the sum of the utilities of the flows of packets at the different nodes. For the first algorithm, the paper proves rate stability for strictly feasible arrival rates and also Harris recurrence of the queues. For the second algorithm, the paper proves the $\epsilon$-optimality. Both algorithms operate with strictly *local* information in the case of decreasing step sizes, and operate with the additional information of the number of nodes in the network in the case of constant step size.

## 1 Introduction

The problem of scheduling and controlling congestion in networks with conflicting nodes has received considerable attention over the last few years for communication networks, stochastic processing networks (cf. [22], [21]) and switched networks (cf. [50]).

Chronologically, the major steps are efficient random access algorithms, the stability of maximum weight scheduling (MW), randomized versions of MW, greedy algorithms with good throughput properties, and optimal local algorithms.

A number of efficient random access algorithms for scheduling transmissions of nodes were proposed, starting with the classical ALOHA protocol [1, 38]. Hajek and van Loon [20] first showed that an adaptive version of ALOHA achieves the maximum throughput possible for that network. Works by Kelly and McPhee [29, 28, 35], Mosely and Humblet [43], Tsybakov and

---

[*]L. Jiang and J. Walrand are with the department of EECS at University of California, Berkeley. D. Shah and J. Shin are with the departments of EECS and Mathematics respectively, at Massachusetts Institute of Technology. The author list is in the alphabetical order. This work was supported in parts by NSF projects HSD 0729361, CNS 0546590, TF 0728554, DARPA ITMANET project and MURI grant BAA 07-036.18. Authors' email addresses are: {ljiang, wlr}@eecs.berkeley.edu, {devavrat, jinwoos}@mit.edu.

Likhanov [55], Aldous [2], Hastad, Leighton and Rogoff [23], Goldberg et al. [17] establish various negative and positive results about the setup when time is slotted, packets are unit size and packets may be queued or not queued. These papers assume that the nodes do not sense the transmission of other nodes. For an online survey (until October 2002) of contention resolution without carrier sense, see [18]. More recently, Gupta and Stolyar [19] and Stolyar [52] proposed algorithms that can achieve the capacity of slotted ALOHA by dynamically adjusting the access probabilities. Another class of random access algorithm is CSMA (Carrier Sense Multiple Access). For example, Eryilmaz, Marbach and Ozdaglar [36] showed that with a particular interference model ("primary interference model"), properly choosing the access probabilites in CSMA can achieve the maximum throughput in the asymptotic regime of small sensing delay and large networks. A related work by Bordenave, McDonald and Proutiére [4] analyzes the 'capacity' of large network (or mean field limit) for a given set of access probabilities.

The MW algorithm was proposed by Tassiulas and Ephremides [54]. This algorithm schedules the independent set (non-conflicting nodes) with the maximum sum of queue lengths. These authors show that the sum of the squares of the queue lengths is a Lyapunov function, thus proving stability. Variants of this algorithm have good delay properties (cf. Shah and Wischik [49, 50]). Unfortunately, finding the MW independent set is NP-complete, making such algorithms difficult to implement. The central idea of considering the maximization of the sum of the user utilities is due to [30]. See also [34, 40]. Combining this objective with the scheduling appears in [44, 45] as well as [12, 13]. For a related survey, see [8, 51].

Randomized versions of MW by Tassiulas [53] and its variant by Giaccone, Prabhakar and Shah [16] provide a simpler (centralized) implementation of MW for input-queued switches while retaining the throughput property. A distributed implementation of this algorithm based on distributed sampling and distributed (a la gossip, cf. Shah [48]) summation procedure was proposed by Modiano, Shah and Zussman [41]. This algorithm, though simple and distributed, requires network-wide information exchange for each new scheduling decision. To overcome this limitation, Rajagopalan, Shah and Shin [46] proposed a distributed, simple, throughput optimal algorithm that is *Markovian* in which each node exchanges exactly *one* message/number (through broadcast transmission) with its neighbor at each time. This algorithm tries to design a reversible Markovian algorithm based on a Metropolis and Hasting's method that solves a network-wide optimization problem inspired by MW algorithm. The choice of weight of a queue as an appropriate function of the queue-size plays key role in establishing the throughput optimality. Authors conjecture that a simplified version of their algorithm that performs *no* information exchange is throughput optimal. The conjecture, as this paper is written, remains unresolved. An interested reader can find a summary of design and analysis of MW-based scheduling algorithms (till 2007) for switched networks in a book chapter by Shah [47].

Greedy algorithms are simpler than MW. Parallel Iterative Matching [3] and iSLIP [37] were shown to be 50% throughput optimal [9]. Subsequently, Dimakis and Walrand [11] identified sufficient conditions on the network topology for throughput optimality. Those conditions were further weakened to obtain fractional throughput results about a class of wireless networks by Joo, Lin and Shroff [27] and Leconte, Ni and Srikant [31]. These algorithms are generally not throughput optimal and require multiple rounds of message exchanges among nodes.

A class of local algorithms was proposed by Jiang and Walrand [26]. The algorithms adjust access probabilities in CSMA for both scheduling and congestion control by means of a novel optimization problem and its relation to certain reversible networks. The result is a totally distributed algorithm. They conjecture it to be throughput optimal and utility maximizing for scheduling and congestion control respectively. In [25], the authors use a suggestion by Shah to adjust the rates over increasing intervals and they adapt techniques from stochastic approximation to prove the convergence, rate stability, and optimality of the algorithms in [26]. Independently, Liu et al. [32] showed that, under stringent technical assumptions, the algorithm in [26] converges to an approximate utility maximizing solution. However, their result does not establish the throughput optimality (i.e., stability of queue-size in some form). Further, implicitly their algorithm requires some knowledge about the entire system.

The key idea of [26] is that, instead of using the MW schedule, the algorithm attempts to improve the schedule to match the arrival rates into the queues. The schedule is parameterized by the aggressiveness with which the nodes request the channel, i.e., by a parameter of the backoff time in a CSMA algorithm. One then defines a distance between the actual schedule and the desired schedule. The gradient of that distance with respect to the aggressiveness of one node turns out to be the difference between the average service and arrival rates at that node. Since the queue length reflects this rate difference, the adjustment of the aggressiveness of one node is based on the queue length at that node and is local.

The technical problem to prove the convergence and the optimality of the algorithm in [26] is as follows. The queue length of one node measures the difference between the actual arrivals and services at the node, not between the average values of those quantities, as the algorithm needs. The idea is that over a long enough time, the random quantities approach their mean values. However, the algorithm changes the parameters (the aggressiveness of the nodes). The intuition is that if the parameters remain constant for long enough, then the distribution of the underlying Markov chain approaches its invariant distribution. Consequently, the algorithm based on the queue lengths approaches the desired gradient algorithm. The general idea of using a random version of the desired gradient is at the heart of stochastic approximation (see [5, 6] and [32]). Here, the additional step is to show that the Markov chain approaches its stationary distribution fast enough for the mean values of the observed quantities to be close to the desired gradient. The needed technical tool is a bound on the mixing time of the Markov chain. Here, as in [25], we use a uniformized version of the continuous time Markov chain to exploit a bound available for the mixing time of discrete time Markov chains.

The current paper provides an alternate proof of the rate stability. Moreover, it proves the Harris recurrence of the queue lengths when using a variant of the algorithm that requires that each node knows the total number of nodes in the network. Under that assumption, for any given $\varepsilon > 0$, there is a congestion control algorithm that is $\varepsilon$-optimal . The difference between the proof in [25] and the current proof of rate stability is as follows. In [25], the error between the gradient and its random version is decomposed into a martingale term and a bias. The bias is bounded using the mixing time result. For the martingale term, [25] uses the supermartingale convergence theorem. In the current paper, instead of using the supermatingale convergence, the error enters in a Taylor expansion. The second term of the Taylor expansion involves a Hessian matrix whose entries happen to be correlations that can be bounded, using suitable choices of the adjustment steps in the algorithm. The proof of the Harris recurrence involves constructing a 'petite set.' The intuitive meaning of this set is a generalization of a recurrent state for a countable Markov chain. Here, the state space is not countable but one can find a positive recurrent set of states whose probability transitions are lower bounded by a given measure. Once the Markov chain hits this set, it starts afresh with at least the given measure, thus providing the coupling property that leads to the ergodicity of recurrent chains.

The paper is organized as follows. Section 2 defines the network model. The main results are stated in Section 3. Some preliminaries about Markov chains as well as a relevant (CSMA) Markov chain are introduced in Section 4. The throughput properties of scheduling algorithms are proved in Section 5. Specifically, rate stability and Harris recurrence properties are proved in Section 5.1 and Section 5.2 respectively. Section 6 analyzes the congestion control problem. Section 7 concludes the paper.

## 2 Model and Problem Statement

Our network graph is a collection of $n$ queues. Time is indexed by $t \in \mathbb{R}_+$. Let $Q_i(t) \in \mathbb{R}_+$ denote the amount of work in the $i$th queue at time $t$ and let $\mathbf{Q}(t) = [Q_i(t)]_{1 \leq i \leq n}$. Initially, $t = 0$ and $\mathbf{Q}(0) = \mathbf{0}$, i.e., the system starts empty. Work arrives to each queue either as per an exogenous arrival process or is controlled by each queue as per a certain algorithm. Each queue can potentially be serviced at unit rate resulting in the departure of work from it. Throughout this paper, we shall assume single-hop network. That is, once work departs from a queue, it leaves the network. In this paper, we will not consider multihop network but we believe that the results of this paper can be extended without much difficulty.

The queues are offered service as per the constraint imposed by interference. To define this constraint, let $G = (V, E)$ denote the inference graph between queues. Here vertices $V = \{1, \ldots, n\}$ represent the $n$ queues and edges $E \subset V \times V$ represent interfering queues: $(i, j) \in E$ iff transmissions of queues $i$ and $j$ interfere with each other. Let $\mathcal{N}(i) = \{j \in V : (i, j) \in E\}$ denote the neighbors of node $i$. Let $\sigma_i(t) \in \{0, 1\}$ denote whether queue $i$ is transmitting at time $t$, with notation that $\sigma_i(t) = 1$ represents transmission. Let $\boldsymbol{\sigma}(t) = [\sigma_i(t)]$. Then, interference imposes the constraint that for all $t \in \mathbb{R}_+$,

$$\boldsymbol{\sigma}(t) \in \mathcal{I}(G) \triangleq \{\boldsymbol{\rho} = [\rho_i] \in \{0,1\}^n : \rho_i + \rho_j \leq 1, \ \forall \ (i,j) \in E\}. \tag{1}$$

The resulting queueing dynamics are described as follows. For $0 \leq s < t$ and $1 \leq i \leq n$,

$$Q_i(t) = Q_i(s) - \int_s^t \sigma_i(r) \mathbf{1}_{\{Q_i(r) > 0\}} \, dr + A_i(s, t),$$

where $A_i(s, t)$ denotes the cumulative arrival to queue $i$ in the time interval $(s, t]$ and $\mathbf{1}_{\{\cdot\}}$ denotes the indicator function. Finally, define the cumulative departure process $\boldsymbol{D}(t) = [D_i(t)]$, where

$$D_i(t) = \int_0^t \sigma_i(r) \mathbf{1}_{\{Q_i(r) > 0\}} \, dr.$$

We define the capacity region of such a wireless network. The capacity region $\mathcal{C} \subset [0,1]^n$ is the convex hull of the feasible scheduling set $\mathcal{I}(G)$, i.e.

$$\mathcal{C} = \left\{ \sum_{\boldsymbol{\rho} \in \mathcal{I}(G)} \alpha_{\boldsymbol{\rho}} \boldsymbol{\rho} : \sum_{\boldsymbol{\rho} \in \mathcal{I}(G)} \alpha_{\boldsymbol{\rho}} = 1, \text{ and } \alpha_{\boldsymbol{\rho}} \geq 0 \text{ for all } \boldsymbol{\rho} \in \mathcal{I}(G) \right\}.$$

The intuition behind this definition of capacity region comes from the fact that any algorithm has to choose a schedule from $\mathcal{I}(G)$ at each time and hence the time average of the 'service rate' induced by any algorithm must belong to $\mathcal{C}$.

**Scheduling Problem.** In this setup, we assume that the arrival process at each queue is exogeneous. Recall that $A_i(s,t)$ denotes the work that has arrived to queue $i$ in the time interval $(s,t]$; $A_i(t) \triangleq A_i(0,t)$ represents cumulative arrival process. We assume that the increments in the arrival process over integral times, i.e., $A_i(k, k+1)$ for $k \in \mathbb{Z}_+$, are independent and identically distributed with bounded support. Moreover, we assume that $A_i(1) \in [0, K]$ and $\Pr(A_i(1) = 0) > 0$ for all $i$. Note that this setup naturally allows for $A_i$ and $A_j$ to be very different processes for $i \neq j$. Finally, we define $\lambda_i = \mathbb{E}[A_i(1)]$. Under our setup the strong law of large numbers implies that

$$\lim_{t \to \infty} \frac{A_i(t)}{t} = \lambda_i, \text{ with probability } 1. \quad (2)$$

Let $\boldsymbol{\lambda} = [\lambda_i]$. We assume that $\lambda_{\min} \triangleq \min_{1 \leq i \leq n} \lambda_i > 0$ without loss of generality[1]. In this setup, we need a scheduling algorithm that decides $\boldsymbol{\sigma}(t)$ each time instant $t \in \mathbb{R}_+$. Intuitively, we would expect that a good algorithm will keep the queues as small as possible. To make this notion formal, first note that if $\boldsymbol{\lambda} \notin \boldsymbol{\Lambda}$, then no algorithm can keep the queues finite, where

$$\boldsymbol{\Lambda} = \left\{ \boldsymbol{\lambda} \in \mathbb{R}_+^n : \boldsymbol{\lambda} \leq \gamma \text{ componentwise, for some } \gamma \in \mathcal{C} \right\}.$$

Motivated by this observation, we call $\boldsymbol{\lambda}$ *strictly admissible* if $\boldsymbol{\lambda} \in \boldsymbol{\Lambda}^o$, where

$$\boldsymbol{\Lambda}^o = \left\{ \boldsymbol{\lambda} \in \mathbb{R}_+^n : \boldsymbol{\lambda} < \gamma \text{ componentwise, for some } \gamma \in \mathcal{C} \right\}.$$

We call a scheduling algorithm *rate stable* if for any $\boldsymbol{\lambda} \in \boldsymbol{\Lambda}^o$, the following holds with probability 1:

$$\lim_{t \to \infty} \frac{1}{t} D_i(t) = \lambda_i \quad \forall \ 1 \leq i \leq n.$$

Given (2), this is equivalent to

$$\lim_{t \to \infty} \frac{Q_i(t)}{t} = 0, \quad \forall \ 1 \leq i \leq n.$$

---

[1] Note that, if $\lambda_i = 0$ for some $i$, then algorithm will ignore such queues by setting their access probability to 0.

Rate stability is a weaker notion of *throughput optimality* or *stability* of the network. A stronger notion requires that for any $\boldsymbol{\lambda} \in \boldsymbol{\Lambda}^o$ the underlying network Markov process is *positive recurrent* or more generally *positive Harris recurrent*.

In summary, the problem of scheduling requires designing an algorithm that makes the network-wide decisions $\sigma(t) \in \mathcal{I}(G)$ for all $t$ so that the network is throughput optimal (rate stable or positive recurrent). The algorithm should utilize only local information, i.e., $\sigma_i(t)$ should be based on the *history* observed at node $i$ only and the *sensing* information available at node $i$ about which of its neighbors are transmitting at a given time.

**Congestion Control Problem.** In this setup, unlike the scheduling problem, we require each node or queue to control its arrival or data generation process. Specifically, at each node $i$, an algorithm decides the rate $\lambda_i(t) \in [0, 1]$ at each time $t$. The data is generated at node $i$ as per a deterministic process with rate $\lambda_i(t)$ at time $t$. That is, for any $0 \leq s < t$,

$$A_i(s,t) = \int_s^t \lambda_i(r) dr.$$

Given the arrival or data generation process, the remaining problem is similar to scheduling. That is, an algorithm is required to make decisions $\sigma(t) \in \mathcal{I}(G)$ for all $t$ using only local information and so as to keep queues small, if possible. Now in order to determine the right rate allocation, we assume that all nodes have some utility. Let $U_i : [0,1] \to \mathbb{R}$ be a strictly concave and increasing utility function of node $i$, with $U_i(x)$ representing the value of its utility when it is allocated rate $x \in [0,1]$. Then, ideally we wish nodes to allocate rates $\boldsymbol{\lambda}^* = [\lambda_i^*]$ where

$$\boldsymbol{\lambda}^* \;=\; \mathsf{argmax} \;\; \sum_{i=1}^n U_i(\lambda_i) \quad \text{over} \quad \boldsymbol{\lambda} \in \boldsymbol{\Lambda}. \tag{3}$$

In summary, the problem of congestion control requires designing an algorithm that makes decisions $\boldsymbol{\lambda}(t) \in [0,1]^n$ and $\sigma(t) \in \mathcal{I}(G)$ for all $t$ so that $\boldsymbol{\lambda}(t) \to \boldsymbol{\lambda}^*$ and the network of queues is stable, i.e. rate stable or more generally positive recurrent. The algorithm should utilize only local information, i.e., both $\lambda_i(t)$ and $\sigma_i(t)$ should be based on the *history* observed at node $i$ only and the *sensing* information available at node $i$ about which of its neighbors are transmitting at a given time.

## 3 Main Results

This section describes our algorithms and theorems stating their performance guarantees for scheduling and congestion control. The algorithms presented here are variants of algorithms proposed in an earlier work [26]. As noted earlier, this paper provides an alternate proof of the rate stability established in [25] and the new result of Harris recurrence.

### 3.1 Scheduling Algorithm

The algorithm to decide $\sigma(t)$ through local decisions $\sigma_i(t)$ can be classified as a CSMA (carrier sense random access) algorithm. The basic operation of each node under such an algorithm can be described as follows. In between two transmissions, a node waits for a random amount of time – also known as *backoff*. Each node can sense the medium perfectly and instantly, i.e., knows if any other interferring node is transmitting at a given time instance. If a node that finishes waiting senses the medium to be busy, it starts waiting for another random amount of time; else, it starts transmitting for a random amount of time. The nodes repeat this operation. The difference between all such protocols lies in the selection of the random waiting time and random transmission time.

In this paper, we assume that node $i$'s random waiting time and transmission time have exponential distributions with mean $1/R_i$ and 1, respectively. Therefore, the performance of algorithm is solely determined by the parameters $R_i, 1 \leq i \leq n$. In essence, our scheduling algorithm will *learn* a good value for $R_i$ at each node $i$ using only local information, so that the performance of the algorithm is throughput optimal. It is somewhat surprising that such a simple class of algorithms can indeed achieve the optimal throughput.

More precisely, let $R_i(t)$ be the value of parameter $R_i$ at time $t$. Given that $R_i(t)$ changes over time, the waiting time becomes distributed according to an exponential distribution with time varying rate. A convenient way to think of this is as follows. Suppose node $i$ starts its new waiting period at time $t_1$ and is still waiting at time $t > t_1$. Then, given the history till time $t$, the waiting time ends during $(t, t+\varepsilon)$ with probability $R_i(t)\varepsilon + o(\varepsilon)$.

Given the above description, the scheduling algorithm is completely determined once we describe how $R_i(t)$ are decided for all $i$ and all $t \in \mathbb{R}_+$. For convenience, we describe the algorithm for selecting $r_i(t) \triangleq \ln R_i(t)$. The algorithm, at each node $i$, updates $r_i(t)$ at time instances $L(j), j \in \mathbb{Z}_+$ with $L(0) = 0$. Also, $r_i(t)$ remains the same between times $L(j)$ and $L(j+1))$ for all $j \in \mathbb{Z}_+$. To begin with, the algorithm sets $r_i(0) = 0$ for all $i$. With an abuse of notation, from now onwards we denote by $r_i(j)$ the value of $r_i(t)$ for all $t \in [L(j), L(j+1))$. Finally, define $T(j) = L(j+1) - L(j)$ for $j \geq 0$. Note that $T(0) = L(1) - L(0) = L(1)$.

In what follows, we describe two variants that differ in the choice of $L(j)$ and the update procedure $r_i(\cdot)$. The first variant uses strictly local information while the second variant uses information about the number of nodes in the network and a performance parameter $\varepsilon > 0$. We provide theorems quantifying the performance of these variants as well.

**Scheduling Algorithm 1.** In this variant, we use a varying update interval $T(j)$. Specifically, we select

$$T(j) = \exp\left(\sqrt{j}\right), \quad \text{for} \ \ j \geq 1.$$

Also, we choose a step-size $\alpha(j)$ of the algorithm as

$$\alpha(j) = \frac{1}{j}, \quad \text{for} \ \ j \geq 1.$$

Given this, node $i$ updates $r_i(\cdot)$ as follows. Let $\widehat{\lambda}_i(j), \widehat{s}_i(j)$ be empirical arrival and service observed at queue $i$ in $[L(j), L(j+1))$. That is,

$$\widehat{\lambda}_i(j) = \frac{1}{T(j)} A_i(L(j), L(j+1)), \quad \text{and} \ \ \widehat{s}_i(t) = \frac{1}{T(j)} \left[\int_{L(j)}^{L(j+1)} \sigma_i(t)dt\right].$$

Then, the update $r_i(j+1)$ of $r_i(j)$ is defined by

$$r_i(j+1) = r_i(j) + \alpha(j)(\widehat{\lambda}_i(j) - \widehat{s}_i(j)), \qquad (4)$$

with initial condition $r_i(0) = 0$. This update rule is essentially an approximate gradient algorithm for the optimization problem (26) defined below.

Note that, under this update rule, the algorithm at each node $i$ uses only its local history. Despite this, we establish that this algorithm is rate stabile. Formally, we obtain the following result.

**Theorem 1** *The scheduling algorithm with updating rule (4) as described above is rate stable for any $\boldsymbol{\lambda} \in \Lambda^o$.*

**Scheduling Algorithm 2.** In this variant, we use $T(j) = T$ for some fixed $T$. The choice of $T$ will be depend on two quantities – the number $n$ of nodes in the network (we assume $n > 3$ here) and $\varepsilon > 0$ that characterizes the approximate stability of the system. Specifically,

$$T \triangleq T(n, \varepsilon) = \exp\left(\Theta\left(\frac{n^2}{\varepsilon} \log \frac{n}{\varepsilon}\right)\right). \qquad (5)$$

Then the updating rule becomes

$$r_i(j+1) = \left[r_i(j) + \alpha(\widehat{\lambda}_i(j) + \varepsilon - \widehat{s}_i(j))\right]_{\frac{n}{\varepsilon}}, \qquad (6)$$

where $\alpha = \alpha(n, \varepsilon) = \varepsilon^2 n^{-2}/72(K+1)^2$ (here, $K$ is the Lipschitz constant for cumulative arrival process) and if $\widehat{\mathbf{x}} = [\mathbf{x}]_{\frac{n}{\varepsilon}}$ then

$$\widehat{x}_i = \begin{cases} \frac{n}{\varepsilon} & \text{if } x_i > \frac{n}{\varepsilon} \\ -\frac{n}{\varepsilon} & \text{if } x_i < -\frac{n}{\varepsilon} \\ x_i & \text{otherwise.} \end{cases} \qquad (7)$$

We state the following throughput optimal property of the algorithm using this rule.

**Theorem 2** *For given $\varepsilon > 0$, under the above described scheduling algorithm the network is positive Harris recurrent if $\boldsymbol{\lambda} + \mathbf{1} \cdot 2\varepsilon \in \Lambda^o$.*

## 3.2 Congestion Control Algorithm

The algorithm for congestion control has to select the appropriate values of $r_i(\cdot)$ and the arrival rates $\lambda_i(\cdot)$. These decisions have to be taken so that the arrival rates maximize overall network utility while keeping the queues small.

Like in the scheduling problem, the algorithm for congestion control updates its choice of $r_i(t)$ and $\lambda_i(t)$ at time instances $L(j), j \in \mathbb{Z}_+$ with $L(0) = 0$. To begin with, it sets $r_i(0) = 0$ and $\lambda_i(0) = 1$ for all $i$. With an abuse of notation, from now onwards we denote by $r_i(j)$ (resp. $\lambda_i(j)$) the value of $r_i(t)$ (resp. $\lambda_i(t)$) for all $t \in [L(j), L(j+1))$. As before, define $T(j) = L(j+1) - L(j)$ for $j \geq 0$. Note that $T(0) = L(1) - L(0) = L(1)$.

In what follows, we describe two algorithms for congestion control. Like the two scheduling algorithms, the first variant does not utilize any global information while the second variant utilizes information about number of nodes and a performance parameter.

**Congestion Control Algorithm 1.** Here, $T(j) = \exp(\sqrt{j})$, $\alpha(j) = \frac{1}{j}$ for $j \in \mathbb{N}$. The $r_i(\cdot), \lambda_i(\cdot)$ are updated as follows: for all $i$,

$$\begin{aligned} r_i(j+1) &= [r_i(j) + \alpha(j)(\lambda_i(j) - \widehat{s}_i(j))]_+, \\ \lambda_i(j+1) &= \arg\max_{y \in [0,1]} \left(\beta \cdot U_i(y) - r_i(j+1)y\right), \end{aligned} \quad (8)$$

with initially $\mathbf{r}(0) = \mathbf{0}$ and $\boldsymbol{\lambda}(0) = \mathbf{1}$. Here $\beta > 0$ is an algorithm parameter and it plays a role in determining the efficiency of the algorithm. As before, each node updates its parameters based only on local information. Recall that, each node $i$ accepts data at rate $\lambda_i(j)$ in $[L(j), L(j+1))$ deterministically. We state the following result about the performance of this algorithm.

**Theorem 3** *Under the above described algorithm, the queues $\mathbf{Q}(\cdot)$ and arrival rates $\boldsymbol{\lambda}(\cdot)$ are such that*

$$\lim_{t \to \infty} \frac{Q_i(t)}{t} = 0, \quad \text{and} \quad \lim_{j \to \infty} \boldsymbol{\lambda}(j) = \bar{\boldsymbol{\lambda}}, \quad \text{with probability 1,}$$

*where $\bar{\boldsymbol{\lambda}}$ is such that*

$$\left(\sum_i U_i(\bar{\lambda}_i)\right) \geq \left(\sum_i U_i(\lambda_i^*)\right) - \frac{\log |\mathcal{I}(G)|}{\beta}, \quad (9)$$

*where recall that $\boldsymbol{\lambda}^*$ is a solution to utility maximization problem (3).*

**Congestion Control Algorithm 2.** Here, the step-size $T(j)$ is constant, and equal a large value $T$, for all $j$. In addition to the above, we assume that $U_i(\cdot)$ are such that

$$V = \max_i U_i'(0) < \infty, \quad (10)$$

and $V$ is known to all nodes. The algorithm performance parameter is $\varepsilon > 0$. The step size $\alpha$ is a small, fixed constant in $(0,1)$. Let $\beta = 4n/\varepsilon$. Select $T$ such that

$$T = \exp\left(\Theta(\beta n V)\right) \Theta\left(\frac{(\beta V + \alpha)n^2}{\beta \varepsilon}\right).$$

The updating rule is as follows. For all $i$,

$$\begin{aligned} r_i(j+1) &= [r_i(j) - \alpha \widehat{s}_i(j)]_+ + \alpha \lambda_i(j), \\ \lambda_i(j+1) &= \arg\max_{y \in [0,1]} \left(\beta \cdot U_i(y) - r_i(j+1)y\right), \end{aligned} \quad (11)$$

with initially $\mathbf{r}(0) = \mathbf{0}$ and $\boldsymbol{\lambda}(0) = \mathbf{1}$. We state the following result about this algorithm.

**Theorem 4** *Under the above described algorithm 2, the queue lengths* $\mathbf{Q}(\cdot)$ *are such that*

$$Q_i(t) \leq \frac{T(\beta V + 2\alpha)}{\alpha}, \quad \text{for all } t \geq 0, \quad \text{for all } i.$$

*Further, define* $\tilde{\boldsymbol{\lambda}}(J)$ *as*

$$\tilde{\lambda}_i(J) = \frac{1}{J}\left(\sum_{j=0}^{J-1}\lambda_i(j)\right).$$

*Then,*

$$\liminf_{J\to\infty}\left(\sum_i U_i(\tilde{\lambda}_i(J))\right) \geq \left(\sum_i U_i(\lambda_i^*)\right) - \varepsilon, \qquad \text{with probability } 1. \qquad (12)$$

## 4 Preliminaries

This section recall relevant known results about establishing bound on mixing time of Markov chains. We will start by setting up basic notations and recalling known definitions.

### 4.1 Markov Chain and Mixing Time

Consider a discrete-time, time-homogeneous Markov chain over a finite state space $\Omega$. Let an $|\Omega| \times |\Omega|$ matrix $P$ be its transition probability matrix. If $P$ is irreducible and aperiodic, then the Markov chain has a unique stationary distribution and it is ergodic in the sense that $\lim_{\tau\to\infty} P^\tau(j,i) \to \boldsymbol{\pi}_i$ for any $i,j \in \Omega$. Here $\boldsymbol{\pi} = [\boldsymbol{\pi}_i]$ denotes the stationary distribution of the Markov chain. The adjoint of the transition matrix $P$, also called the time-reversal of $P$, is denoted by $P^*$ and defined as: for any $i, j \in \Omega$, $\boldsymbol{\pi}(i)P^*(i,j) = \boldsymbol{\pi}(j)P(j,i)$. By definition, $P^*$ has $\boldsymbol{\pi}$ as its stationary distribution as well. If $P = P^*$ then $P$ is called *reversible*, and in this paper we will be primarily interested in such reversible Markov chains.

As noted earlier, the distribution of the irreducible and aperiodic Markov chain converges to its stationary distribution $\boldsymbol{\pi}$ starting from any initial condition. To establish our results, we will need quantifiable bounds on the time it takes for the Markov chain to reach close to stationary distribution – popularly known as *mixing time*. To make this notion precise and recall known bound on mixing time, we start with definition of distance between probability distributions.

**Definition 1 (Two distances)** *Given two probability distributions* $\boldsymbol{\mu}$ *and* $\boldsymbol{\nu}$ *on a finite space* $\Omega$, *we define the following two distances. The total variation distance, denoted as* $\|\boldsymbol{\mu} - \boldsymbol{\nu}\|_{TV}$ *is*

$$\|\boldsymbol{\mu} - \boldsymbol{\nu}\|_{TV} = \frac{1}{2}\sum_{i\in\Omega}|\mu_i - \nu_i|.$$

*The* $\chi^2$ *distance, denoted as* $\left\|\frac{\boldsymbol{\nu}}{\boldsymbol{\mu}} - 1\right\|_{2,\boldsymbol{\mu}}$ *is*

$$\left\|\frac{\boldsymbol{\nu}}{\boldsymbol{\mu}} - 1\right\|_{2,\boldsymbol{\mu}}^2 = \|\boldsymbol{\nu} - \boldsymbol{\mu}\|_{2,\frac{1}{\boldsymbol{\mu}}}^2 = \sum_{i\in\Omega}\mu_i\left(\frac{\nu_i}{\mu_i} - 1\right)^2.$$

We make note of the following relation between the above defined two distances: for any probability distributions $\boldsymbol{\mu}, \boldsymbol{\nu}$, using the Cauchy-Schwartz inequality we have

$$\begin{aligned}
\left\|\frac{\boldsymbol{\nu}}{\boldsymbol{\mu}} - 1\right\|_{2,\boldsymbol{\mu}} &= \sqrt{\sum_{i \in \Omega} \mu_i \left(\frac{\nu_i}{\mu_i} - 1\right)^2} \\
&= \sqrt{\sum_{i \in \Omega} \mu_i} \sqrt{\sum_{i \in \Omega} \mu_i \left(\frac{\nu_i}{\mu_i} - 1\right)^2} \\
&\geq \sum_{i \in \Omega} \mu_i \left|\frac{\nu_i}{\mu_i} - 1\right| = \sum_{i \in \Omega} |\nu_i - \mu_i| \\
&= 2 \|\boldsymbol{\nu} - \boldsymbol{\mu}\|_{TV}.
\end{aligned} \qquad (13)$$

In general, for any two vectors $\mathbf{u}, \mathbf{v} \in \mathbb{R}_+^{|\Omega|}$, we define norm

$$\|\mathbf{v}\|_{2,\mathbf{u}}^2 = \sum_{i \in \Omega} u_i v_i^2.$$

This norm naturally induces a matrix norm that will be useful in determining rate of convergence or mixing time of a finite state Markov chain.

**Definition 2 (Matrix norm)** *Consider an $|\Omega| \times |\Omega|$ non-negative valued matrix $A \in \mathbb{R}_+^{|\Omega| \times |\Omega|}$ and a vector $\mathbf{u} \in \mathbb{R}_+^{|\Omega|}$. Then, the matrix norm of $A$ with respect to $\mathbf{u}$ is defined as follows:*

$$\|A\|_\mathbf{u} = \sup_{\mathbf{v}: \mathbb{E}_\mathbf{u}[\mathbf{v}]=0} \frac{\|A\mathbf{v}\|_{2,\mathbf{u}}}{\|\mathbf{v}\|_{2,\mathbf{u}}},$$

where $\mathbb{E}_\mathbf{u}[\mathbf{v}] = \sum_i u_i v_i$.

It can be easily checked that the above definition of matrix norm satisfies the following properties.

**P1**. For matrices $A, B \in \mathbb{R}_+^{|\Omega| \times |\Omega|}$ and $\boldsymbol{\pi} \in \mathbb{R}_+^{|\Omega|}$

$$\|A + B\|_{\boldsymbol{\pi}} \leq \|A\|_{\boldsymbol{\pi}} + \|B\|_{\boldsymbol{\pi}}.$$

**P2**. For matrix $A \in \mathbb{R}_+^{|\Omega| \times |\Omega|}$, $\boldsymbol{\pi} \in \mathbb{R}_+^{|\Omega|}$ and $c \in \mathbb{R}$,

$$\|cA\|_{\boldsymbol{\pi}} = |c| \|A\|_{\boldsymbol{\pi}}.$$

**P3**. Let $A$ and $B$ be transition matrices of reversible Markov chains, i.e. $A = A^*$ and $B = B^*$. Let both of them have $\boldsymbol{\pi}$ as their unique stationary distribution. Then,

$$\|AB\|_{\boldsymbol{\pi}} \leq \|A\|_{\boldsymbol{\pi}} \|B\|_{\boldsymbol{\pi}}.$$

**P4**. Let $A$ be the transition matrix of a reversible Markov chain, i.e. $A = A^*$. Let $\boldsymbol{\pi}$ be its stationary distribution. Then,

$$\|A\|_{\boldsymbol{\pi}} = \lambda_{\max},$$

where $\lambda_{\max} = \max\{|\lambda|| \lambda \neq 1 \text{ is an eigenvalue of } A\}$.

For a probability matrix $P$, mostly in this paper we will be interested in the matrix norm of $P$ with respect to its stationary distribution $\boldsymbol{\pi}$, i.e. $\|P\|_{\boldsymbol{\pi}}$. Therefore, unless stated otherwise if we use matrix norm for a probability matrix without mentioning the reference measure, then it is with respect to the stationary distribution. That is, in above example $\|P\|$ will mean $\|P\|_{\boldsymbol{\pi}}$.

With these definitions and fact that $P$ and $P^*$ have the same stationary distribution, say $\boldsymbol{\pi}$, it follows that for any distribution $\mu$ on $\Omega$

$$\left\| \frac{\mu P}{\boldsymbol{\pi}} - 1 \right\|_{2,\boldsymbol{\pi}} \leq \|P^*\| \left\| \frac{\mu}{\boldsymbol{\pi}} - 1 \right\|_{2,\boldsymbol{\pi}}, \tag{14}$$

where we have used (abused) notaiton $\|P^*\| = \|P^*\|_{\boldsymbol{\pi}}$ and since $\mathbb{E}_{\boldsymbol{\pi}}\left[\frac{\mu}{\boldsymbol{\pi}} - 1\right] = 0$, with interpretation $\frac{\mu}{\boldsymbol{\pi}} = [\mu(i)/\boldsymbol{\pi}(i)]$. Therefore, for a reversible Markov chain ($P = P^*$) starting with initial distribution $\boldsymbol{\mu}(0)$, the distribution $\boldsymbol{\mu}(\tau)$ at time $\tau$ is such that

$$\left\| \frac{\boldsymbol{\mu}(\tau)}{\boldsymbol{\pi}} - 1 \right\|_{2,\boldsymbol{\pi}} \leq \|P\|^{\tau} \left\| \frac{\boldsymbol{\mu}(0)}{\boldsymbol{\pi}} - 1 \right\|_{2,\boldsymbol{\pi}}. \tag{15}$$

Now starting from any state $i$, i.e. probability distribution with unit mass on state $i$, the initial distance $\left\| \frac{\boldsymbol{\mu}(0)}{\boldsymbol{\pi}} - 1 \right\|_{2,\boldsymbol{\pi}}$ in the worst case is bounded above by $\sqrt{1/\boldsymbol{\pi}_{\min}}$ where $\boldsymbol{\pi}_{\min} = \min_i \pi_i$. Therefore, for any $\delta > 0$ we have $\left\| \frac{\mu(\tau)}{\boldsymbol{\pi}} - 1 \right\|_{2,\boldsymbol{\pi}} \leq \delta$ for any $\tau$ such that

$$\tau \geq \frac{\log 1/\boldsymbol{\pi}_{\min} + \log 1/\delta}{\log 1/\|P\|} = O\left( \frac{\log 1/\boldsymbol{\pi}_{\min} + \log 1/\delta}{1 - \|P\|} \right).$$

This suggests that the "mixing time", i.e. time to reach (close to) stationary distribution of the Markov chain scales inversely with $1 - \|P\|$. Therefore, we will define the "mixing time" of a Markov chain with transition matrix $P$ as $1/(1 - \|P\|)$. This also suggests that in order to bound the distance between a Markov chain's distribution after some steps and its stationary distribution, it is sufficient to obtain a bound on $\|P\|$.

## 4.2 CSMA Markov Chain & Its Mixing Time

As the *backbone* of our algorithms, for scheduling and congestion control, is a Markov chain with state space being $\mathcal{I}(G)$. In recent years, this was considered in the context of CSMA by Wang and Kar [56]. Its transition matrix is determined by the vector $\mathbf{r}(\cdot)$ and hence is time varying. However, if $\mathbf{r}(\cdot)$ were fixed, then it will be a time-homogeneous reversible Markov chain. In the context of CSMA, the vector of $\mathbf{r}(\cdot)$ corresponds to the *aggressiveness* of backoff. In what follows, we will describe this time-homogeneous version (i.e. assuming fixed $\mathbf{r}(\cdot)$) of Markov chain, which was implicit in the description of the scheduling/congestion control algorithm, its stationary distribution and a bound on its mixing time.

To this end, let $\mathbf{r}(\cdot) = \mathbf{r}$ be fixed. Recall that, under scheduling/congestion control algorithm, each node does the following. Each node $i$ is either in 'transmission' state (denoted as $\sigma_i = 1$) or 'waiting' state (denoted by $\sigma_i = 0$). In a waiting state, node has an exponential clock ticking at rate $R_i = \exp(r_i)$ (mean $1/R_i$): when it ticks, if medium is free, it acquires and starts transmitting (i.e. now $\sigma_i = 1$); else if medium is busy, it continues the waiting state (i.e. retains $\sigma_i = 0$). In a transmission state, node has an exponential clock ticking at rate 1: when it ticks, it frees the medium and enters waiting state (i.e. now $\sigma_i = 0$).

This is a continuous time Markov chain over a finite state space. It can be easily checked that it has the following *product form* stationary distribution $\boldsymbol{\pi}^{\mathbf{r}} = [\pi^{\mathbf{r}}_{\boldsymbol{\sigma}}]$: for any $\boldsymbol{\sigma} \in \mathcal{I}(G)$,

$$
\begin{aligned}
\pi^{\mathbf{r}}_{\boldsymbol{\sigma}} &\propto \exp(\boldsymbol{\sigma} \cdot \mathbf{r}) \\
&= \frac{\exp(\boldsymbol{\sigma} \cdot \mathbf{r})}{\sum_{\boldsymbol{\sigma} \in \mathcal{I}(G)} \exp(\boldsymbol{\sigma} \cdot \mathbf{r})}.
\end{aligned} \quad (16)
$$

Here, for vectors $\mathbf{a}, \mathbf{b}$, we use notation of dot-product, $\mathbf{a} \cdot \mathbf{b} = \sum_{i=1}^{n} a_i b_i$. Under this stationary distribution the average fraction of time node $i$ ends up transmitting, which is its 'service rate', is given by

$$
s_i(\mathbf{r}) = \mathbb{E}_{\boldsymbol{\pi}^{\mathbf{r}}}[\sigma_i] = \sum_{\boldsymbol{\sigma} \in \mathcal{I}(G)} \sigma_i \cdot \pi^{\mathbf{r}}_{\boldsymbol{\sigma}}. \quad (17)
$$

Throughout, we will call $\mathbf{s}(\mathbf{r}) = [s_i(\mathbf{r})]$ as the service rate vector induced by $\boldsymbol{\pi}^{\mathbf{r}}$. To understand the 'mixing time' of this continous time Markov chain, first consider its following discrete time version with transition matrix $P$ on $\mathcal{I}(G)$. Under $P$, the transition from current state $\boldsymbol{\sigma} \in \mathcal{I}(G)$ to the next state $\boldsymbol{\sigma}^* \in \mathcal{I}(G)$ happens as follows:

- Choose a node $i \in V$ with probability $\frac{\max\{\exp(r_i), 1\}}{\sum_{k \in V} \max\{\exp(r_k), 1\}}$.

- If $\sigma_i = 1$ (equivalently, $i \in \boldsymbol{\sigma}$), then

$$
\sigma^*_i = \begin{cases} 0 & \text{with probability } \min\{1/\exp(r_i), 1\} \\ 1 & \text{otherwise} \end{cases}
$$

and $\sigma^*_j = \sigma_j$ for $j \neq i$.

- If $\sigma_i = 0$ and $\sigma_k = 0$ for all $k \in \mathcal{N}(i)$ (i.e. $i \notin \boldsymbol{\sigma}$ and $k \notin \boldsymbol{\sigma}$ for all $k \in \mathcal{N}(i)$), then

$$
\sigma^*_i = \begin{cases} 1 & \text{with probability } \min\{\exp(r_i), 1\} \\ 0 & \text{otherwise} \end{cases}
$$

and $\sigma^*_j = \sigma_j$ for all $j \neq i$.

- Otherwise $\boldsymbol{\sigma}^* = \boldsymbol{\sigma}$.

The above discrete version of the continuous time Markov chain is reversible, i.e. $P = P^*$. It can be checked that $P$ is indeed the discretized version, i.e. $\boldsymbol{\pi}^{\mathbf{r}}$ is stationary distribution of $P$.

The continuous time Markov chain relates to the above described discrete time Markov chain with transition matrix $P$ as follows: think of continuous time Markov chain making its transitions when a clock of net rate $R = \sum_{k \in V} \max\{\exp(r_k), 1\}$ ticks. And, when its clock ticks the next state for transition is chosen as per transition matrix $P$. Given this, let $\mu(t)$ be the distribution over $\mathcal{I}(G)$ under the continuous CSMA Markov chain at time $t$. Then, the dynamics of $\mu(\cdot)$ is described as

$$\begin{aligned}\boldsymbol{\mu}(t) &= \sum_{i=0}^{\infty} \Pr(\zeta = i)\boldsymbol{\mu}(0)P^i \\ &= \frac{1}{e^{Rt}}\boldsymbol{\mu}(0)e^{RtP} \\ &= \boldsymbol{\mu}(0)e^{Rt(P-I)},\end{aligned} \quad (18)$$

where $\zeta$ is Poisson random variable with parameter $Rt$ which is equal to the number of clock ticks in time $[0,t]$. In the above (and throughout), in the left multiplication of a vector with a matrix, the vector should be thought of as a row vector.

Given (18) and earlier discussion on matrix norms, mixing time analysis for discrete time Markov chain, we obtain that

$$\left\|\frac{\boldsymbol{\mu}(t)}{\boldsymbol{\pi}^{\mathbf{r}}} - 1\right\|_{2,\boldsymbol{\pi}^{\mathbf{r}}} < \left\|e^{Rt(P-I)}\right\| \left\|\frac{\boldsymbol{\mu}(0)}{\boldsymbol{\pi}^{\mathbf{r}}} - 1\right\|_{2,\boldsymbol{\pi}^{\mathbf{r}}}.$$

Therefore, to bound the distance between $\boldsymbol{\mu}(t)$ and $\boldsymbol{\pi}^{\mathbf{r}}$, we need to get a bound on $\left\|e^{Rt(P-I)}\right\|$.

**Lemma 5** *The matrix norm of $e^{Rt(P-I)}$ is bounded as*

$$\left\|e^{Rt(P-I)}\right\| \leq \left(1 - \frac{1}{\exp(\Theta\left(n|\mathbf{r}|_\infty + n\right))}\right)^{\lfloor t \rfloor}.$$

*Proof.* Define partition function or normalization constant $Z(\mathbf{r})$ of $\boldsymbol{\pi}^{\mathbf{r}}$ as

$$Z(\mathbf{r}) = \sum_{\boldsymbol{\sigma} \in \mathcal{I}(G)} \exp(\boldsymbol{\sigma} \cdot \mathbf{r}).$$

It follows that

$$Z(\mathbf{r}) \leq |\mathcal{I}(G)| \exp(n\|\mathbf{r}\|_\infty) \leq \exp(n(1 + \|\mathbf{r}\|_\infty)).$$

Therefore, for any $\boldsymbol{\sigma} \in \mathcal{I}(G)$,

$$\begin{aligned}\pi^{\mathbf{r}}_{\boldsymbol{\sigma}} &= \frac{1}{Z(\mathbf{r})}\exp(\boldsymbol{\sigma} \cdot \mathbf{r}) \\ &\geq \exp(-n(1 + 2\|\mathbf{r}\|_\infty)).\end{aligned} \quad (19)$$

Now for any $\boldsymbol{\sigma}, \boldsymbol{\rho} \in \mathcal{I}(G)$ such that they differ in only one component, i.e. it is possible to transit from $\boldsymbol{\sigma}$ to $\boldsymbol{\rho}$ and vice versa in one-step, we have

$$\begin{aligned}\left(e^{R(P-I)}\right)_{\boldsymbol{\sigma}\boldsymbol{\rho}} &\geq Pr(\zeta = 1)P_{\boldsymbol{\sigma}\boldsymbol{\rho}} \\ &\geq \exp(-\Theta(n\|\mathbf{r}\|_\infty + n)).\end{aligned}$$

In above we used the fact that $\zeta$ is Poisson random variable with parameter $\sum_i \max\{\exp(r_i), 1\}$ which is at most $n(1 + \exp(\|\mathbf{r}\|_\infty))$.

Given above calculations, we are ready to bound the conductance, $\Phi$, of $W = e^{R(P-I)}$ defined as

$$\begin{aligned}
\Phi &= \min_{S \subset \mathcal{I}(G)} \frac{Q(S, \mathcal{I}(G) \setminus S)}{\boldsymbol{\pi}^{\mathbf{r}}(S) \boldsymbol{\pi}^{\mathbf{r}}(\mathcal{I}(G) \setminus S)} \\
&> \min_{\boldsymbol{\sigma}, \boldsymbol{\rho} \in \mathcal{I}(G)} \pi^{\mathbf{r}}_{\boldsymbol{\sigma}} W_{\boldsymbol{\sigma}\boldsymbol{\rho}} \\
&> \exp\left(-\Theta(n\|\mathbf{r}\|_\infty + n)\right),
\end{aligned} \qquad (20)$$

where $Q(A, B) = \sum_{\boldsymbol{\sigma} \in A, \boldsymbol{\rho} \in B} \pi^{\mathbf{r}}_{\boldsymbol{\sigma}} W_{\boldsymbol{\sigma}\boldsymbol{\rho}}$. By Cheeger's inequality [24, 10, 33, 42], it is well known that

$$\begin{aligned}
\lambda_{\max} &\leq 1 - \frac{\Phi^2}{2} \\
&< 1 - \exp\left(-\Theta(n|r|_{\max} + n)\right).
\end{aligned} \qquad (21)$$

where $|r|_{\max} := \|\mathbf{r}\|_\infty$.

Hence, from the properties **P3** and **P4** of the matrix norm, we can conclude that

$$\begin{aligned}
\|e^{Rt(P-I)}\| &\leq \|e^{R(P-I)}\|^{\lfloor t \rfloor} \\
&\leq \lambda_{\max}^{\lfloor t \rfloor} \\
&< (1 - \exp(-\Theta(n|r|_{\max} + n)))^{\lfloor t \rfloor}.
\end{aligned} \qquad (22)$$

$\square$

Using Lemma 5 and the fact that $\left\|\frac{\boldsymbol{\mu}(0)}{\boldsymbol{\pi}^{\mathbf{r}}} - 1\right\|_{2,\boldsymbol{\pi}} < \sqrt{1/\pi_{\min}} < \exp(\Theta(n|\mathbf{r}|_\infty + n))$, we obtain

$$\left\|\frac{\boldsymbol{\mu}(t)}{\boldsymbol{\pi}^{\mathbf{r}}} - 1\right\|_{2,\boldsymbol{\pi}^{\mathbf{r}}} < \delta \quad \text{for} \quad t > \exp(\Theta(n\|\mathbf{r}\|_\infty + n)) \log \frac{1}{\delta}. \qquad (23)$$

### 4.3 Positive Harris Recurrence

We recall the well known notion of positive Harris recurrence for discrete time Markov chains. To this end, consider a time homogeneous discrete time Markov chain over a polish space $\mathsf{X}$ denoted as $X(\tau) \in \mathsf{X}$ for time $\tau \in \mathbb{N} \cup \{0\}$. Let $\mathcal{B}_\mathsf{X}$ be the Borel $\sigma$-algebra of $\mathsf{X}$ with respect to this product topology. Let $P$ denote the probability transition matrix of this discrete-time $\mathsf{X}$-valued Markov chain. Given a probability distribution (also called sampling distribution) $a$ on $\mathbb{N}$, the $a$-sampled transition matrix of the Markov chain, denoted by $K_a$ is defined as

$$K_a(\mathbf{x}, B) = \sum_{\tau \geq 0} a(\tau) P^\tau(\mathbf{x}, B), \quad \text{for any} \ \ \mathbf{x} \in \mathsf{X}, \ B \in \mathcal{B}_\mathsf{X}.$$

Now we recall notion of a *petite* set [39]. A non-empty set $A \in \mathcal{B}_\mathsf{X}$ is called $\mu_a$-*petite* if $\mu_a$ is a non-trivial measure on $(\mathsf{X}, \mathcal{B}_\mathsf{X})$ and $a$ is a probability distribution on $\mathbb{N}$ such that for any $\mathbf{x} \in A$,

$$K_a(\mathbf{x}, \cdot) \geq \mu_a(\cdot).$$

A set is called a *petite* set if it is $\mu_a$-petite for some such non-trivial measure $\mu_a$.

We will call the Markov chain positive Harris recurrence if there exists a closed petite set that is positive recurrent. This is formally summarized as the following known result (see book by Meyn and Tweedie [39] or survey by Foss and Konstantopoulos [14] for details).

**Theorem 6** *Let $B$ be a closed petite set. Suppose $B$ is recurrent, i.e.*

$$\Pr(T_B < \infty | X(0) = \mathbf{x}) = 1, \text{ for any } \mathbf{x} \in \mathsf{X},$$

*where $T_B = \inf\{\tau \geq 1 : X(\tau) \in B\}$. Further, let*

$$\sup_{\mathbf{x} \in B} \mathbb{E}_{\mathbf{x}}[T_B] < \infty.$$

*Then the Markov chain is positive Harris recurrent.*

Theorem 6 suggests that to establish the positive Harris recurrence of the network Markov chain, it is sufficient to find a closed petite set that satisfies the conditions of Theorem 6. To establish recurrence property of a set, the following Lyapunov and Foster's criteria will be useful.

**Lemma 7** *Let there exist functions $h, g : \mathsf{X} \to \mathbb{R}$ and $L : \mathsf{X} \to \mathbb{R}_+$ such that for any $\mathbf{x} \in \mathsf{X}$,*

$$\mathbb{E}\left[L(X(g(\mathbf{x}))) - L(X(0)) | X(0) = \mathbf{x}\right] \leq -h(\mathbf{x}),$$

*and*

*(a) $\inf_{\mathbf{x} \in \mathsf{X}} h(\mathbf{x}) > -\infty$,*

*(b) $\liminf_{L(\mathbf{x}) \to \infty} h(\mathbf{x}) > 0$,*

*(c) $\sup_{L(\mathbf{x}) \leq \gamma} g(\mathbf{x}) < \infty$ for all $\gamma > 0$,*

*(d) $\limsup_{L(\mathbf{x}) \to \infty} g(\mathbf{x})/h(\mathbf{x}) < \infty$.*

*Then, there exists finite $\kappa > 0$ so that the set $B_\kappa = \{\mathbf{x} : L(\mathbf{x}) \leq \kappa\}$, the following holds:*

$$\mathbb{E}_{\mathbf{x}}[T_{B_\kappa}] < \infty, \quad \text{for any } \mathbf{x} \in \mathsf{X}$$
$$\sup_{\mathbf{x} \in B_\kappa} \mathbb{E}_{\mathbf{x}}[T_{B_\kappa}] < \infty.$$

## 5 Throughput Property of Scheduling Algorithms

This section establish throughput optimality for the two scheduling algorithm proposed in Section 3.1. Specifically, we present proof of Theorem 1 to establish rate stability of the scheduling algorithm 1 in Section 5.1 and proof of Theorem 2 to establish positive Harris recurrence of the scheduling algorithm 2 in Section 5.2. As noted earlier, the algorithm 1 does not utilize any global information while the algorithm 2 utilizes only global information in terms of number of nodes in the network.

## 5.1 Proof of Theorem 1: Rate stability

The proof of Theorem 1 consists of the three parts. First, we introduce and study a relevant optimization problem whose parameters are the vector of backoff parameters $\mathbf{r}(\cdot)$. On one hand, it is related to the classical variational principle studied in the context of Gibbs distributions or Markov Random Fields (e.g. Chapter 15.4, Georgii[15]). On the other hand, it will suggest that the optimal solution corresponding to $\mathbf{r}(\cdot)$, say $\mathbf{r}^*$, will be such that the service rate vector $\mathbf{s}(\mathbf{r}^*)$, induced by the Markov chain's stationary distribution, is the same as the arrival rate vector $\boldsymbol{\lambda}$. Therefore, if Algorithm 1 adjusts the $\mathbf{r}(\cdot)$ appropriately so that $\mathbf{r}(\cdot)$ converges to $\mathbf{r}^*$, then there is a possibility establishing rate stability. In the second part, we do so by showing that the Algorithm 1 is a stochastic gradient algorithm for the optimization problem of interest. Finally, in the third part, we conclude the proof of Theorem 1 by establishing that the system is rate stable for any $\boldsymbol{\lambda} \in \boldsymbol{\Lambda}^o$.

**A relevant optimization problem & its properties.** We begin by introducing the optimization problem of interest. Its relation to variational principle will be alluded later. To this, given an arrival rate vector $\boldsymbol{\lambda} \in \boldsymbol{\Lambda}^o$ and $\mathbf{r} \in \mathbb{R}^n$, define function $F(\mathbf{r}, \boldsymbol{\lambda})$, where

$$F(\mathbf{r}, \boldsymbol{\lambda}) = \boldsymbol{\lambda} \cdot \mathbf{r} - \log\left(\sum_{\boldsymbol{\sigma} \in \mathcal{I}(G)} \exp(\boldsymbol{\sigma} \cdot \mathbf{r})\right). \tag{24}$$

The interpretation of $F(\mathbf{r}, \boldsymbol{\lambda})$ is as follows. Assume that $\boldsymbol{\lambda}$ is strictly feasible, i.e. $\boldsymbol{\lambda} \in \boldsymbol{\Lambda}^o$, so that it can be written as a positive combination of feasible transmission vectors. That is,

$$\boldsymbol{\lambda} = \sum_{\boldsymbol{\sigma} \in \mathcal{I}(G)} \nu_{\boldsymbol{\sigma}} \boldsymbol{\sigma}$$

for $\boldsymbol{\nu} = [\nu_{\boldsymbol{\sigma}}] \in \mathbb{R}_+^{|\mathcal{I}(G)|}$. Therefore, if $\boldsymbol{\sigma} \in \mathcal{I}(G)$ is scheduled for $\nu_{\boldsymbol{\sigma}}$ fraction of the time then effective service rate is the same as arrive rate $\boldsymbol{\lambda}$. Clearly, $\boldsymbol{\nu}$ can be thought of as a probability distribution on $\mathcal{I}(G)$ as well.

Now consider the Kullback-Liebler divergence or relative entropy between this distribution $\boldsymbol{\nu}$ and $\boldsymbol{\pi}^{\mathbf{r}}$, the stationary distribution of CSMA Markov chain with parameters $\mathbf{r}$, defined as follows:

$$d(\boldsymbol{\nu}, \boldsymbol{\pi}^{\mathbf{r}}) = \sum_{\boldsymbol{\sigma} \in \mathcal{I}(G)} \nu_{\boldsymbol{\sigma}} \log\left(\frac{\nu_{\boldsymbol{\sigma}}}{\pi_{\boldsymbol{\sigma}}^{\mathbf{r}}}\right).$$

It is well known that

$$\|\boldsymbol{\nu} - \boldsymbol{\pi}^{\mathbf{r}}\|_{TV} \leq d(\boldsymbol{\nu}, \boldsymbol{\pi}^{\mathbf{r}}).$$

However, $d(\cdot, \cdot)$ is not a metric and it's only pre-metric. Consider the following relation between $F(\mathbf{r}, \boldsymbol{\lambda})$ and $d(\boldsymbol{\nu}, \boldsymbol{\pi}^{\mathbf{r}})$:

$$F(\mathbf{r}, \boldsymbol{\lambda}) = \left(\sum_{\boldsymbol{\rho} \in \mathcal{I}(G)} \nu_{\boldsymbol{\rho}} \boldsymbol{\rho} \cdot \mathbf{r}\right) - \log\left(\sum_{\boldsymbol{\sigma} \in \mathcal{I}(G)} \exp(\boldsymbol{\sigma} \cdot \mathbf{r})\right)$$

$$= \sum_{\boldsymbol{\rho} \in \mathcal{I}(G)} \nu_{\boldsymbol{\rho}} \log\left(\frac{\exp(\boldsymbol{\rho} \cdot \mathbf{r})}{\sum_{\boldsymbol{\sigma} \in \mathcal{I}(G)} \exp(\boldsymbol{\sigma} \cdot \mathbf{r})}\right)$$

$$\begin{aligned}
&= \sum_{\rho \in \mathcal{I}(G)} \nu_\rho \log\left(\frac{\pi_\rho^{\mathbf{r}}}{\nu_\rho}\right) + \sum_{\rho \in \mathcal{I}(G)} \nu_\rho \log \nu_\rho \\
&= -d(\boldsymbol{\nu}, \boldsymbol{\pi}^{\mathbf{r}}) - H_{ER}(\boldsymbol{\nu}).
\end{aligned} \qquad (25)$$

Thus, for a given fixed $\boldsymbol{\lambda}$, we have that

$$d(\boldsymbol{\nu}, \boldsymbol{\pi}^{\mathbf{r}}) + F(\mathbf{r}, \boldsymbol{\lambda}) = \text{Constant}.$$

Therefore, minimizing $d(\boldsymbol{\nu}, \boldsymbol{\pi}^{\mathbf{r}})$ with respect to parameter $\mathbf{r}$ is equivalent to maximizing $F(\mathbf{r}, \boldsymbol{\lambda})$. And as we shall show, that this optimization of $\mathbf{r}$ leads to $\mathbf{r}^*$ so that the $\mathbf{s}(\mathbf{r}^*)$ equals $\boldsymbol{\lambda}$ as long as $\boldsymbol{\lambda} \in \Lambda^o$. For this reason, the following is the optimization problem of interest.

$$\begin{aligned}
\text{maximize} \quad & F(\mathbf{r}, \boldsymbol{\lambda}) \\
\text{subject to} \quad & \mathbf{r} \in \mathbb{R}^n.
\end{aligned} \qquad (26)$$

Now we state the following useful properties of this optimization problem.

**Lemma 8** *Consider a given $\boldsymbol{\lambda} \in \mathbb{R}_+^n$. Then, the following holds.*

(1) *The objective function $F(\mathbf{r}, \boldsymbol{\lambda})$, as a function of $\mathbf{r}$ is strictly concave. Moreover,*

$$\frac{d}{dr_i} F(\mathbf{r}, \boldsymbol{\lambda}) = \lambda_i - s_i(\mathbf{r}) \qquad (27)$$

*and*

$$\frac{\partial^2 F}{\partial r_i \partial r_j} = \mathbb{E}_{\boldsymbol{\pi}^{\mathbf{r}}}[\sigma_i \sigma_j] - \mathbb{E}_{\boldsymbol{\pi}^{\mathbf{r}}}[\sigma_i]\mathbb{E}_{\boldsymbol{\pi}^{\mathbf{r}}}[\sigma_j]. \qquad (28)$$

(2) *For $\boldsymbol{\lambda} \in \Lambda^o$, the optimization problem (26) has a unique solution $\mathbf{r}^*(\boldsymbol{\lambda})$ that is attained and $F(\mathbf{r}^*(\boldsymbol{\lambda})) < 0$. Let $\mathbf{r} = \mathbf{r}^*(\boldsymbol{\lambda})$. Then, under $\boldsymbol{\pi}^{\mathbf{r}}$ the 'service rate vector' (as defined in (17)) $\mathbf{s}(\mathbf{r}) = [s_i(\mathbf{r})]$ equals $\boldsymbol{\lambda}$. That is,*

$$\sum_{\boldsymbol{\sigma} \in \mathcal{I}(G)} \sigma_i \pi_{\boldsymbol{\sigma}}^{\mathbf{r}} = \lambda_i, \quad \text{for all } i.$$

(3) *Further, for any $\varepsilon > 0$ such that $\boldsymbol{\lambda} + \varepsilon \mathbf{1} \in \Lambda$,*

$$|\mathbf{r}^*(\boldsymbol{\lambda})|_\infty \leq \frac{\log |\mathcal{I}(G)|}{\min\{\varepsilon, \boldsymbol{\lambda}_{\min}\}}.$$

*Proof.* For simplicity of notation, we will drop the reference to $\boldsymbol{\lambda}$ in $F(\mathbf{r}, \boldsymbol{\lambda})$ and simply denote it as $F(\mathbf{r})$ as we have $\boldsymbol{\lambda}$ fixed throughout the proof. We will use additional notation of the partition function $Z(\mathbf{r})$ of $\boldsymbol{\pi}^{\mathbf{r}}$ defined as

$$Z(\mathbf{r}) = \sum_{\boldsymbol{\sigma} \in \mathcal{I}(G)} \exp(\boldsymbol{\sigma} \cdot \mathbf{r}).$$

*Proof of (1).* We wish to establish that $F(\mathbf{r})$ is strictly concave as a function of $\mathbf{r}$. To this end, its first derivative can be calculated as,

$$\begin{aligned}
\frac{\partial F}{\partial r_i} &= \lambda_i - \frac{\sum_{\boldsymbol{\sigma} \in \mathcal{I}(G)} \sigma_i \exp(\boldsymbol{\sigma} \cdot \mathbf{r})}{\sum_{\boldsymbol{\rho} \in \mathcal{I}(G)} \exp(\boldsymbol{\rho} \cdot \mathbf{r})} \\
&= \lambda_i - \sum_{\boldsymbol{\sigma} \in \mathcal{I}(G)} \sigma_i \pi_{\boldsymbol{\sigma}}^{\mathbf{r}} \\
&= \lambda_i - \mathbb{E}_{\boldsymbol{\pi}^{\mathbf{r}}}[\sigma_i] \;=\; \lambda_i - s_i(\mathbf{r}).
\end{aligned} \qquad (29)$$

Here, we have used the definition of $\boldsymbol{\pi}^{\mathbf{r}}$ in (16).

To obtain strict concavity, we would like to show that the Hessian of $F$ is negative definite. Now, we compute the second derivative as (using (29))

$$\begin{aligned}
-\frac{\partial^2 F}{\partial r_i \partial r_j} &= \frac{\partial}{\partial r_j} \mathbb{E}_{\boldsymbol{\pi}^{\mathbf{r}}}[\sigma_i] = \frac{\partial}{\partial r_j} \left( \sum_{\boldsymbol{\sigma} \in \mathcal{I}(G)} \sigma_i \exp(\boldsymbol{\sigma} \cdot \mathbf{r}) \frac{1}{Z(\mathbf{r})} \right) \\
&= \sum_{\boldsymbol{\sigma} \in \mathcal{I}(G)} \left( \sigma_i \sigma_j \exp(\boldsymbol{\sigma} \cdot \mathbf{r}) \frac{1}{Z(\mathbf{r})} \right) + \sum_{\boldsymbol{\sigma} \in \mathcal{I}(G)} \left( \sigma_i \exp(\boldsymbol{\sigma} \cdot \mathbf{r}) \frac{\partial}{\partial r_j} \frac{1}{Z(\mathbf{r})} \right) \\
&= \mathbb{E}_{\boldsymbol{\pi}^{\mathbf{r}}}[\sigma_i \sigma_j] - \sum_{\boldsymbol{\sigma} \in \mathcal{I}(G)} \sigma_i \exp(\boldsymbol{\sigma} \cdot \mathbf{r}) \frac{1}{Z(\mathbf{r})^2} \left( \sum_{\boldsymbol{\rho} \in \mathcal{I}(G)} \rho_j \exp(\boldsymbol{\rho} \cdot \mathbf{r}) \right) \\
&= \mathbb{E}_{\boldsymbol{\pi}^{\mathbf{r}}}[\sigma_i \sigma_j] - \left( \sum_{\boldsymbol{\sigma} \in \mathcal{I}(G)} \sigma_i \frac{\exp(\boldsymbol{\sigma} \cdot \mathbf{r})}{Z(\mathbf{r})} \right) \times \left( \sum_{\boldsymbol{\rho} \in \mathcal{I}(G)} \rho_j \frac{\exp(\boldsymbol{\rho} \cdot \mathbf{r})}{Z(\mathbf{r})} \right) \\
&= \mathbb{E}_{\boldsymbol{\pi}^{\mathbf{r}}}[\sigma_i \sigma_j] - \mathbb{E}_{\boldsymbol{\pi}^{\mathbf{r}}}[\sigma_i] \mathbb{E}_{\boldsymbol{\pi}^{\mathbf{r}}}[\sigma_j].
\end{aligned} \qquad (30)$$

Thus, the Hessian of $F$, denoted by $M = [M_{ij}]$ with $M_{ij} = \frac{\partial^2 F}{\partial r_i \partial r_j}$, is the negative covariance matrix of a random vector with distribution $\boldsymbol{\pi}^{\mathbf{r}}$. It is well known that covariance matrices are positive semi-definite, i.e., $M$ is negative semi-definite. For strict concavity of $F$, we need to show that $M$ is negative definite or the covariance matrix of $\boldsymbol{\pi}^{\mathbf{r}}$ is positive definite. To this end, let $\mathbf{X}$ be a vector (of $n$ binary) random variables with the joint distribution $\boldsymbol{\pi}^{\mathbf{r}}$. Let $\boldsymbol{\mu} = \mathbb{E}[\mathbf{X}] \in \mathbb{R}_+^n$ be the vector of its mean. Then, from the above we have that $-M = \mathbb{E}[(\mathbf{X} - \boldsymbol{\mu})(\mathbf{X} - \boldsymbol{\mu})^T]$. Now consider any vector $\zeta \in \mathbb{R}^n$. To establish the positive definiteness of $-M$, we need to show that

$$\zeta^T (-M) \zeta > 0 \Leftrightarrow \zeta \neq \mathbf{0}.$$

Suppose to the contrary that there exists a vector $\zeta \neq \mathbf{0}$ such that $\zeta^T(-M)\zeta = 0$. Clearly,

$$\zeta^T(-M)\zeta = \zeta^T \mathbb{E}[(\mathbf{X} - \boldsymbol{\mu})(\mathbf{X} - \boldsymbol{\mu})^T] \zeta = \mathbb{E}[\zeta^T(\mathbf{X} - \boldsymbol{\mu})(\mathbf{X} - \boldsymbol{\mu})^T \zeta] \geq 0.$$

Therefore, let us assume that

$$\zeta^T(-M)\zeta = \mathbb{E}[\zeta^T(\mathbf{X} - \boldsymbol{\mu})(\mathbf{X} - \boldsymbol{\mu})^T \zeta] = 0.$$

That is, the random variable $\zeta^T(\mathbf{X} - \boldsymbol{\mu}) = 0$ with probability 1 with respect to $\boldsymbol{\pi}^{\mathbf{r}}$. Now consider $n$ vectors $e_1, \ldots, e_n$, where in $e_i$ only node $i$ is selected; i.e., $e_i \in \{0, 1\}^n$ with $i$th component 1 and all other components 0. Now, by definition $\pi^{\mathbf{r}}(e_i) > 0$ for any $\mathbf{r}$. Therefore, the above condition implies that for all $i$, $\zeta^T(e_i - \boldsymbol{\mu}) = 0$. That is, for all $i$,

$$\zeta_i(1 - \mu_i) - \sum_{j \neq i} \zeta_j \mu_j = 0 \quad \Rightarrow \quad \zeta_i = \sum_{j=1}^n \zeta_j \mu_j. \tag{31}$$

That is, for all $i$, $\zeta_i = c$. Now applying the same argument with the choice of $\boldsymbol{\sigma} = \mathbf{0}$, we obtain that

$$c \mathbf{1}^T \boldsymbol{\mu} = 0.$$

This immediately implies that $c = 0$ since $\boldsymbol{\mu}^T \mathbf{1} > 0$ for any $\mathbf{r}$. Thus, we have proved that if $\zeta^T(-M)\zeta \leq 0$ then it must be that $\zeta = \mathbf{0}$. That is, $M$ is negative definite and hence $F$ is strictly concave. This completes the proof of (1) of Lemma 8.

*Proof of (2) and (3).* We wish to establish that for $\boldsymbol{\lambda} \in \boldsymbol{\Lambda}^o$, the optimization problem has a unique solution that is attained. We will establish this by showing that the optimal solution must lie inside a closed, bounded and convex set since $\boldsymbol{\lambda} \in \boldsymbol{\Lambda}^o$. As a byproduct, this will provide (3). Then, the strict concavity of $F$ will immediately lead to the existence of a unique solution, and the claim that $\boldsymbol{\lambda} = \mathbf{s}(\mathbf{r}^*(\boldsymbol{\lambda}))$ as a result of the local optimality condition. As the first step towards this, we establish that $F(\mathbf{r}^*(\boldsymbol{\lambda})) < 0$.

To this end, since $\boldsymbol{\lambda} \in \boldsymbol{\Lambda}^o$, it can be easily check that there exists a distribution $\boldsymbol{\nu}$ on $\mathcal{I}(G)$ such that

$$\boldsymbol{\lambda} = \sum_{\boldsymbol{\sigma} \in \mathcal{I}(G)} \nu_{\boldsymbol{\sigma}} \boldsymbol{\sigma}. \tag{32}$$

Therefore, using (32) in the definition of $F$, we have

$$\begin{aligned} F(\mathbf{r}) &= \left( \sum_{\boldsymbol{\rho} \in \mathcal{I}(G)} \nu_{\boldsymbol{\rho}} \boldsymbol{\rho} \cdot \mathbf{r} \right) - \log \left( \sum_{\boldsymbol{\sigma} \in \mathcal{I}(G)} \exp(\boldsymbol{\sigma} \cdot \mathbf{r}) \right) \\ &= \left( \sum_{\boldsymbol{\rho} \in \mathcal{I}(G)} \nu_{\boldsymbol{\rho}} \log \exp(\boldsymbol{\rho} \cdot \mathbf{r}) \right) - \log \left( \sum_{\boldsymbol{\sigma} \in \mathcal{I}(G)} \exp(\boldsymbol{\sigma} \cdot \mathbf{r}) \right) \\ &= \sum_{\boldsymbol{\rho} \in \mathcal{I}(G)} \nu_{\boldsymbol{\rho}} \log \left( \frac{\exp(\boldsymbol{\rho} \cdot \mathbf{r})}{\sum_{\boldsymbol{\sigma} \in \mathcal{I}(G)} \exp(\boldsymbol{\sigma} \cdot \mathbf{r})} \right) \\ &< 0, \end{aligned} \tag{33}$$

The last step follows because (i) for any $\boldsymbol{\rho} \in \mathcal{I}(G)$, $\exp(\boldsymbol{\rho} \cdot \mathbf{r}) < Z(\mathbf{r})$ since any graph has at least two independent sets; (ii) for some $\boldsymbol{\rho} \in \mathcal{I}(G)$, $\nu_{\boldsymbol{\rho}} > 0$.

Next, we will show that if $\boldsymbol{\lambda} + \varepsilon \mathbf{1} \in \boldsymbol{\Lambda}$, then

$$\sup_{\mathbf{r} \in \mathbb{R}^n} F(\mathbf{r}) = \sup_{\mathbf{r} \in [-K, K]^n} F(\mathbf{r}), \quad \text{where} \quad K = \frac{\log |\mathcal{I}(G)|}{\min\{\varepsilon, \boldsymbol{\lambda}_{\min}\}}. \tag{34}$$

To establish (34), we will show that for any $\mathbf{r} \in \mathbb{R}^n$, if (a) $\mathbf{r}_{\max} := \max_{1 \leq i \leq n} r_i > K \geq \log |\mathcal{I}(G)|/\varepsilon$ or (b) $\mathbf{r}_{\min} := \min_{1 \leq i \leq n} r_i < -K$ then $F(\mathbf{r}) \leq F(\mathbf{0}) = -\log |\mathcal{I}(G)|$. As a byproduct, this will imply (3) of Lemma 8.

First for case (a), consider a given $\mathbf{r}$ so that $\mathbf{r}_{\max} > \log |\mathcal{I}(G)|/\varepsilon$. Since $\boldsymbol{\lambda} + \varepsilon \cdot \mathbf{1} \in \Lambda$ and $\boldsymbol{\sigma} \in \{0,1\}^n$, there exists a non-negative valued measure $\boldsymbol{\nu}$ on $\mathcal{I}(G)$ such that

$$\boldsymbol{\lambda} = \sum_{\boldsymbol{\sigma} \in \mathcal{I}(G)} \nu_{\boldsymbol{\sigma}} \boldsymbol{\sigma}, \quad \text{and} \quad \sum_{\boldsymbol{\sigma} \in \mathcal{I}(G)} \nu_{\boldsymbol{\sigma}} \leq 1 - \varepsilon. \tag{35}$$

This implies the existence of a distribution $\widehat{\boldsymbol{\nu}}$ on $\mathcal{I}(G)$ defined as

$$\widehat{\nu}_{\boldsymbol{\sigma}} = \begin{cases} \nu_{\boldsymbol{\sigma}} + \left(1 - \sum_{\boldsymbol{\rho} \in \mathcal{I}(G)} \nu_{\boldsymbol{\rho}}\right) & \text{if } \boldsymbol{\sigma} = \mathbf{0} \\ \nu_{\boldsymbol{\sigma}} & \text{otherwise.} \end{cases}$$

Note that $\boldsymbol{\lambda} = \sum_{\boldsymbol{\sigma} \in \mathcal{I}(G)} \widehat{\nu}_{\boldsymbol{\sigma}} \boldsymbol{\sigma}$. Therefore,

$$\begin{aligned}
F(\mathbf{r}) &= \boldsymbol{\lambda} \cdot \mathbf{r} - \log \left( \sum_{\boldsymbol{\sigma} \in \mathcal{I}(G)} \exp(\boldsymbol{\sigma} \cdot \mathbf{r}) \right) \\
&= \sum_{\boldsymbol{\rho} \in \mathcal{I}(G)} \widehat{\nu}_{\boldsymbol{\rho}} \log \frac{\exp(\boldsymbol{\rho} \cdot \mathbf{r})}{\sum_{\boldsymbol{\sigma} \in \mathcal{I}(G)} \exp(\boldsymbol{\sigma} \cdot \mathbf{r})} \\
&\leq \widehat{\nu}_{\mathbf{0}} \log \frac{\exp(\mathbf{0} \cdot \mathbf{r})}{\sum_{\boldsymbol{\sigma} \in \mathcal{I}(G)} \exp(\boldsymbol{\sigma} \cdot \mathbf{r})} \\
&\leq \varepsilon \log \frac{1}{\exp(\mathbf{r}_{\max})} \\
&< -\log |\mathcal{I}(G)| \\
&= F(\mathbf{0}) \\
&\leq \sup_{\mathbf{r}} F(\mathbf{r}). \tag{36}
\end{aligned}$$

Now, we prove case (b). For this, let $\mathbf{r}$ be such that $\mathbf{r}_{\min} < -\log |\mathcal{I}(G)|/\min\{\varepsilon, \boldsymbol{\lambda}_{\min}\}$. Let $i$ be such that $r_i = \mathbf{r}_{\min}$. Define $\bar{\boldsymbol{\lambda}}$ as $\bar{\lambda}_i = \lambda_i - \min\{\varepsilon, \lambda_i\}$ and $\bar{\lambda}_j = \lambda_j$ for $j \neq i$. Clearly, $\bar{\boldsymbol{\lambda}} + \min\{\varepsilon, \lambda_i\} \cdot \mathbf{1} \in \Lambda^o$. Therefore, similar to (35), there exists non-negative valued measure $\boldsymbol{\nu}^1$ on $\mathcal{I}(G)$ so that

$$\bar{\boldsymbol{\lambda}} = \sum_{\boldsymbol{\sigma} \in \mathcal{I}(G)} \nu^1_{\boldsymbol{\sigma}} \boldsymbol{\sigma}, \quad \text{and} \quad \sum_{\boldsymbol{\sigma} \in \mathcal{I}(G)} \nu^1_{\boldsymbol{\sigma}} \leq 1 - \min\{\varepsilon, \lambda_i\}. \tag{37}$$

Now define a distribution $\boldsymbol{\nu}'$ on $\mathcal{I}(G)$ such that

$$\nu'_{\boldsymbol{\sigma}} = \begin{cases} \nu^1_{\boldsymbol{\sigma}} + \min\{\varepsilon, \lambda_i\} & \text{if } \boldsymbol{\sigma} = e_i, \\ \nu^1_{\boldsymbol{\sigma}} + (1 - \sum_{\boldsymbol{\rho} \in \mathcal{I}(G)} \nu^1_{\boldsymbol{\rho}}) - \min\{\varepsilon, \lambda_i\}, & \text{if } \boldsymbol{\sigma} = \mathbf{0}, \\ \nu^1_{\boldsymbol{\sigma}} & \text{otherwise.} \end{cases}$$

Here, as before, $e_i$ refers to the independent set with only node $i$ transmitting. Note that $\boldsymbol{\lambda} = \sum_{\boldsymbol{\sigma} \in \mathcal{I}(G)} \nu'_{\boldsymbol{\sigma}} \boldsymbol{\sigma}$. Now, combined with the fact that $r_i < -\log|\mathcal{I}(G)|/\min\{\varepsilon, \boldsymbol{\lambda}_{\min}\}$, we have

$$\begin{aligned}
F(\mathbf{r}) &= \boldsymbol{\lambda} \cdot \mathbf{r} - \log\left(\sum_{\boldsymbol{\sigma} \in \mathcal{I}(G)} \exp(\boldsymbol{\sigma} \cdot \mathbf{r})\right) \\
&= \sum_{\boldsymbol{\rho} \in \mathcal{I}(G)} \nu'_{\boldsymbol{\rho}} \log \frac{\exp(\boldsymbol{\rho} \cdot \mathbf{r})}{\sum_{\boldsymbol{\sigma} \in \mathcal{I}(G)} \exp(\boldsymbol{\sigma} \cdot \mathbf{r})} \\
&\leq \nu'_{e_i} \log \frac{\exp(e_i \cdot \mathbf{r})}{\sum_{\boldsymbol{\sigma} \in \mathcal{I}(G)} \exp(\boldsymbol{\sigma} \cdot \mathbf{r})} \\
&\leq \min\{\varepsilon, \boldsymbol{\lambda}_{\min}\} \log \frac{\exp(r_i)}{\exp(0)} \\
&< -\log|\mathcal{I}(G)| \\
&= F(\mathbf{0}) \\
&\leq \sup_{\mathbf{r}} F(\mathbf{r}).
\end{aligned}$$

This completes the proof of (b), and subsequently that of Lemma 8.

**Convergence of $\mathbf{r}(j)$ to $\mathbf{r}^*(\boldsymbol{\lambda})$.** The statement of Lemma 8 suggests that if indeed we have algorithm parameter $\mathbf{r}(j) = \mathbf{r}^*(\boldsymbol{\lambda})$, then we have a desirable situation where the effective service rate equals the arrival rate for all nodes as long as $\boldsymbol{\lambda} \in \Lambda^o$. To this end, we establish that indeed $\mathbf{r}(j)$ converges to $\mathbf{r}^*(\boldsymbol{\lambda})$ with probability 1. And this is because update (4) of scheduling algorithm 1 is essentially step of an approximate gradient algorithm for solving optimization problem (26). This is made precise in the proof of the following Lemma.

**Lemma 9** *If $\boldsymbol{\lambda} \in \Lambda^o$, then under scheduling algorithm 1,*

$$\lim_{j \to \infty} \mathbf{r}(j) = \mathbf{r}^*(\boldsymbol{\lambda}), \quad \text{component-wise, with probability 1.}$$

*Proof.* First note that, the solution $\mathbf{r}^*(\boldsymbol{\lambda})$ of concave (maximization) optimization problem (26) can be found iteratively using the gradient algorithm with appropriate step size. The objective is $F(\mathbf{r}, \boldsymbol{\lambda})$ – we will drop reference to $\boldsymbol{\lambda}$ since it is fixed in what follows and use $F(\mathbf{r})$ instead for $F(\mathbf{r}, \boldsymbol{\lambda})$. Now the $i$th component of gradient vector of $F(\mathbf{r})$, $\nabla F(\mathbf{r})$ is

$$\frac{\partial F}{\partial r_i}(\mathbf{r}) = \lambda_i - s_i(\mathbf{r}).$$

For a given $i$, as per (4) the $r_i(\cdot)$ is updated as

$$\begin{aligned}
r_i(j+1) &= r_i(j) + \alpha(j)\left(\widehat{\lambda}_i(j) - \widehat{s}_i(j)\right) \\
&= r_i(j) + \frac{1}{j}(\lambda_i - s_i(\mathbf{r}(j)) + e(j)) \\
&= r_i(j) + \frac{1}{j}\left(\frac{\partial F_1}{\partial r_i}(\mathbf{r}(j)) + e(j)\right),
\end{aligned}$$

where $e(j) = (\widehat{\lambda}_i(j) - \widehat{s}_i(j)) - (\lambda_i - s_i(\mathbf{r}(j)))$ captures the 'approximation' error in estimating the actual gradient direction given by $\lambda_i - s_i(\mathbf{r}(j))$. Thus, if $e(j) = 0$ then the update of $\mathbf{r}(j)$ is as per the standard gradient algorithm with step size $\alpha(j) = 1/j$. Then standard arguments from optimization theory would imply that $\mathbf{r}(j) \to \mathbf{r}^*(\lambda)$. But, $e(j)$ is a random vector. Therefore, in order to establish the convergence, we will show that norm of $e(j)$ is sufficiently small enough. Specifically, we establish the following.

**Lemma 10** *The following bound holds:*

$$\mathbb{E}\left[\|e(j)\|_1\right] \leq \mathbb{E}\left[\|\widehat{\boldsymbol{\lambda}}(j) - \boldsymbol{\lambda}\|_1 + \|\widehat{\mathbf{s}}(j) - \mathbf{s}(\mathbf{r}(j))\|_1\right] = O\left(\frac{n}{j^2}\right), \tag{38}$$

*where constant in O-term in the error may depend on n.*

The proof of Lemma 10 is stated in Section 5.1.1. Now using the bound of (38) we will establish the convergence of $\mathbf{r}(j) \to \mathbf{r}^*(\boldsymbol{\lambda})$. To this end, consider evolution of $F(\mathbf{r}(\cdot))$. By Taylor's expansion (with notation $\boldsymbol{\delta}(j) = \nabla F(\mathbf{r}(j)) + e(j)$),

$$\begin{aligned}
F(\mathbf{r}(j+1)) &= F\left(\mathbf{r}(j) + \frac{1}{j}[\nabla F(\mathbf{r}(j)) + e(j)]\right) \\
&= F(\mathbf{r}(j)) + \nabla F(\mathbf{r}(j)) \cdot \frac{1}{j}\boldsymbol{\delta}(j) + \frac{1}{2j^2}\boldsymbol{\delta}(t)^T M \boldsymbol{\delta}(j) \\
&\geq F(\mathbf{r}(j)) + \frac{1}{j}\|\nabla F(\mathbf{r}(j))\|_2^2 + \frac{1}{j}\nabla F(\mathbf{r}(j)) \cdot e(j) + \frac{1}{2j^2}\boldsymbol{\delta}(t)^T M \boldsymbol{\delta}(j) \\
&\geq F(\mathbf{r}(j)) + \frac{1}{j}\|\nabla F(\mathbf{r}(j))\|_2^2 - \frac{\|\nabla F(\mathbf{r}(j))\|_\infty \|e(j)\|_1}{j} - \frac{\|M\|_\infty \|\boldsymbol{\delta}(j)\|_1^2}{2j^2}. \end{aligned} \tag{39}$$

Here $M$ is an $n \times n$ matrix as per Taylor's expansion is evaluation of 2nd order partial derivates of $F$ at some values. Therefore any element of $M$, say $M_{ab}$ with $1 \leq a, b \leq n$, is bounded as (using calculations executed in (30))

$$\begin{aligned}
|M_{ab}| &\leq \sup_{\mathbf{r}} \left|\frac{\partial^2 F}{\partial r_a \partial r_b}(\mathbf{r})\right| \\
&= \sup_{\mathbf{r}} |\mathbb{E}_{\boldsymbol{\pi}^{\mathbf{r}}}[\sigma_a]\mathbb{E}_{\boldsymbol{\pi}^{\mathbf{r}}}[\sigma_b] - \mathbb{E}_{\boldsymbol{\pi}^{\mathbf{r}}}[\sigma_a \sigma_b]| \\
&\leq 1. \end{aligned} \tag{40}$$

We also note each component of vectors $\nabla F(\mathbf{r}(j))$ and $e(j)$ are bounded by a constant since the cumulative arrival process is Lipschitz and service process is bounded above by unit rate. Specifically, for any $j$

$$\|\nabla F(\mathbf{r}(j))\|_\infty \leq 2, \quad \|e(j)\|_\infty \leq K+1 \;\Rightarrow\; \|\boldsymbol{\delta}(j)\|_\infty \leq K+3. \tag{41}$$

Taking expection on both sides of (39) and using (40), (41) and Lemma 10, for all $j \geq C$

$$\begin{aligned}
\mathbb{E}\left[F(\mathbf{r}(j+1))\right] &\geq \mathbb{E}\left[F(\mathbf{r}(j))\right] + \frac{1}{j}\mathbb{E}\left[\|\nabla F(\mathbf{r}(j))\|_2^2\right] - O\left(\mathbb{E}[\frac{\|e(j)\|_1}{j}]\right) - O\left(\frac{n^2}{j^2}\right) \\
&\geq \mathbb{E}\left[F(\mathbf{r}(j))\right] + \frac{1}{j}\mathbb{E}\left[\|\nabla F(\mathbf{r}(j))\|_2^2\right] - O\left(\frac{n^2}{j^2}\right). \end{aligned} \tag{42}$$

Performing summation of (42) from $j = C$ to $\infty$, we obtain

$$\sum_{j=C}^{\infty} \frac{1}{j} \mathbb{E}\left[\|\nabla F(\mathbf{r}(j))\|_2^2\right] \leq O(n^2) - \mathbb{E}[F(\mathbf{r}(C))] + F(\mathbf{r}^*(\boldsymbol{\lambda}))$$
$$< \infty, \tag{43}$$

since $F(\mathbf{r}^*(\boldsymbol{\lambda})) < 0$ from Lemma 8 and by definition of the algorithm and $F(\cdot)$, $\mathbb{E}\left[F_1(\mathbf{r}(C))\right] > -\infty$. Now since $\sum_{j=C}^{\infty} 1/j = \infty$, we conclude from (43) that

$$0 = \liminf_j \mathbb{E}\left[\|\nabla F(\mathbf{r}(j))\|_2^2\right]$$
$$\geq \mathbb{E}\left[\liminf_j \|\nabla F(\mathbf{r}(j))\|_2^2\right], \quad \text{by Fatou's Lemma.} \tag{44}$$

Therefore, using property of concave maximization we have that with probability 1,

$$\liminf_j \|\nabla F(\mathbf{r}(j))\|_2 = 0 \;\Rightarrow\; \liminf_j \|\mathbf{r}(j) - \mathbf{r}^*(\boldsymbol{\lambda})\|_2 = 0. \tag{45}$$

Thus, in order to complete the proof of Lemma 9, it is enough to show that $\|\mathbf{r}(j) - \mathbf{r}^*(\boldsymbol{\lambda})\|_2$ converges with probability 1. To this end, consider (with notation $\mathbf{r}^* = \mathbf{r}^*(\boldsymbol{\lambda})$, $\boldsymbol{\delta}(j) = \nabla F(\mathbf{r}(j)) + e(j)$)

$$\begin{aligned}
\|\mathbf{r}(j+1) - \mathbf{r}^*\|_2^2 &= \left\|(\mathbf{r}(j) - \mathbf{r}^*) + \frac{1}{j}\left(\nabla F(\mathbf{r}(j)) + e(j)\right)\right\|_2^2 \\
&= \|\mathbf{r}(j) - \mathbf{r}^*\|_2^2 + \frac{\|\nabla F(\mathbf{r}(j)) + e(j)\|_2^2}{j^2} + \frac{2\boldsymbol{\delta}(j) \cdot (\mathbf{r}(j) - \mathbf{r}^*)}{j} \\
&\overset{(a)}{\leq} \|\mathbf{r}(j) - \mathbf{r}^*\|_2^2 + O\left(\frac{1}{j^2}\right) + \frac{2\nabla F(\mathbf{r}(j)) \cdot (\mathbf{r}(j) - \mathbf{r}^*) + 2e(j) \cdot (\mathbf{r}(j) - \mathbf{r}^*)}{j} \\
&\overset{(b)}{\leq} \|\mathbf{r}(j) - \mathbf{r}^*\|_2^2 + O\left(\frac{1}{j^2}\right) + \frac{2e(j) \cdot (\mathbf{r}(j) - \mathbf{r}^*)}{j} \\
&\overset{(c)}{\leq} \|\mathbf{r}(j) - \mathbf{r}^*\|_2^2 + O\left(\frac{1}{j^2}\right) + O\left(\frac{(\log j + |\mathbf{r}^*|_\infty)\|e(j)\|_1}{j}\right) \\
&\overset{(d)}{\leq} \|\mathbf{r}(j) - \mathbf{r}^*\|_2^2 + O\left(\frac{1}{j^2}\right) + O\left(\frac{\log j}{j}\|e(j)\|_1\right).
\end{aligned}$$

In above, (a) follows from (41), (b) follows from the concavity of $F$, i.e. $\nabla F(\mathbf{r}(j)) \cdot (\mathbf{r}(j) - \mathbf{r}^*) \leq 0$, (c) follows from property of update rule that $|\mathbf{r}(j)|_\infty = O(\log j)$ and (d) from Lemma 8 that $\|\mathbf{r}^*\|_\infty = O(1)$. An application of Lemma 10, we have that

$$\sum_{j=1}^{\infty} \mathbb{E}\left[\|e(j)\|_1\right] < \infty.$$

Since the terms in above are non-negative, by an application of Fubini's theorem and Markov's inequality, we have that with probability 1

$$\sum_{j=1}^{\infty} \|e(j)\|_1 < \infty.$$

Of course, $\sum_j 1/j^2$ is finite. Using this, we have that

$$\|\mathbf{r}(j+1) - \mathbf{r}^*\|_2^2 \leq \|\mathbf{r}(j) - \mathbf{r}^*\|_2^2 + \gamma_j,$$

where $\sum_j \gamma_j < \infty$ with probability 1. Now the following (standard) fact from analysis (proof is omitted) implies that $\|\mathbf{r}(j) - \mathbf{r}^*\|$ convergence with probability 1 and completes the proof of Lemma 9.

**Proposition 11** *Consider two real valued sequences $x_k, y_k, k \in \mathbb{N}$ such that for each $k$,*

$$x_{k+1} \leq x_k + y_k, \quad \text{and} \quad \sum_{k=1}^{\infty} y_k < \infty.$$

*Then, $\lim_k x_k$ exists.*

**Wrapping up: establishing rate stability.** As an implicatin of Lemma 9, we establish the rate stability of the queueing network. The following Lemma implies Theorem 1.

**Lemma 12** *Given $\boldsymbol{\lambda} \in \Lambda^o$, under scheduling algorithm 1,*

$$\lim_{t \to \infty} \frac{Q_i(t)}{t} = 0, \quad \text{for all} \ \ 1 \leq i \leq n.$$

*Proof.* Given $\boldsymbol{\lambda} \in \Lambda^o$, recall that $\mathbf{r}^*(\boldsymbol{\lambda})$ is the unique optimal solution of optimization problem (26) as per Lemma 8. In the remainder of the proof, since $\boldsymbol{\lambda}$ is fixed, we will use notation $F(\mathbf{r}) = F(\mathbf{r}, \boldsymbol{\lambda})$, and $\mathbf{r}^* = \mathbf{r}^*(\boldsymbol{\lambda})$ as before. Now by Lemma 9, we have $\mathbf{r}(j) \to \mathbf{r}^*(\boldsymbol{\lambda})$ with probability 1 as $j \to \infty$. Now as noted earlier, $\nabla F(\mathbf{r}) = \boldsymbol{\lambda} - \mathbf{s}(\mathbf{r})$. It can be easily checked that $\mathbf{s}(\mathbf{r})$ is continuous as function of $\mathbf{r}$. Therefore with probability 1,

$$\lim_{j \to \infty} \nabla F(\mathbf{r}(j)) = \nabla F(\mathbf{r}^*)$$
$$= \mathbf{0}, \tag{46}$$

where the equality to $\mathbf{0}$, the vector of all 0s, is implied by Lemma 8. Thus, effectively

$$\lim_{j \to \infty} \mathbf{s}(\mathbf{r}(j)) = \boldsymbol{\lambda}. \tag{47}$$

Lemma 10 implies that with probability 1,

$$\sum_{j=C}^{\infty} \|\widehat{\boldsymbol{\lambda}}(j) - \boldsymbol{\lambda}\|_1 + \sum_{j=C}^{\infty} \|\widehat{\mathbf{s}}(j) - \mathbf{s}(\mathbf{r}(j))\|_1 < \infty. \tag{48}$$

That is, with probability 1,

$$\lim_{j \to \infty} \widehat{\boldsymbol{\lambda}}(j) = \boldsymbol{\lambda} \quad \text{and} \quad \lim_{j \to \infty} \|\widehat{\mathbf{s}}(j) - \mathbf{s}(\mathbf{r}(j))\| = 0. \tag{49}$$

From (47) and (49), with probability 1,

$$\lim_{j \to \infty} \|\widehat{\boldsymbol{\lambda}}(j) - \widehat{\mathbf{s}}(j)\| = 0. \tag{50}$$

Now consider a node $i$ and any time $t$. Let $t \in [L(j), L(j+1)] = [L(j), L(j) + T(j)]$ for some $j$. We will bound $Q_i(t)/t$ next. To begin with, note that

$$A_i(0,t) = \sum_{k=0}^{j-2} T(k)\widehat{\lambda}_i(k) + A_i(L(j-1), t).$$

Note that the service provide to the $i$th node in interval $[L(k), L(k+1)]$ is $T(k)\widehat{s}_i(k)$. Now, for the purpose of upper bounding queue, we will assume that this service can be used only to serve the work that has arrived in interval $[L(k-1), L(k)]$. Given this, we obtain the following upper bound (using $Q_i(0) = 0$):

$$\begin{aligned} Q_i(t) &= A_i(0,t) - \int_0^t \sigma_i(r) \mathbf{1}_{\{Q_i(r) > 0\}} \, dr \\ &\leq \left[ \sum_{k=0}^{j-2} \left( T(k)\widehat{\lambda}_i(k) - T(k+1)\widehat{s}_i(k+1) \right)_+ \right] + A_i(L(j-1), t). \end{aligned}$$

Here, we have used definition $[x]_+ = \frac{x+|x|}{2}$, the non-negative part of $x$, for any $x \in \mathbb{R}$. Since $t \in [L(j), L(j+1)]$ and the cumulative arrival process is Lipschitz, we have

$$\begin{aligned} A_i(L(j-1), t) &\leq A(L(j-1), L(j+1)) \\ &\leq K(L(j+1) - L(j-1)) \\ &= K(T(j-1) + T(j)). \end{aligned}$$

And, by definition $T(k) \leq T(k+1)$. Therefore, putting these together we obtain

$$\frac{Q_i(t)}{t} \leq \frac{1}{L(j)} \left[ \sum_{k=0}^{j-2} \left( T(k)\widehat{\lambda}_i(k) - T(k)\widehat{s}_i(k+1) \right)_+ \right] + \frac{K(T(j-1) + T(j))}{L(j)}. \quad (51)$$

Consider the first term on the RHS of (51). From (49) and (50), it follows that $\widehat{\lambda}_i(k) - \widehat{s}_i(k+1) \to 0$ as $k \to \infty$. And, $L(j) \geq \sum_{k=0}^{j-2} T(k)$ as well as $L(j) \to \infty$. Therefore, it easily follows that as $j \to \infty$, the first term goes to 0. Now, the second term on the RHS of (51). Since $T(j) = \exp(\sqrt{j})$, $T(j)/L(j) \to 0$ as $j \to \infty$. In summary, from this discussion and (51) we obtain that for any $i$, with probability 1

$$\lim_{t \to \infty} \frac{Q_i(t)}{t} = 0.$$

This complete the proof of Lemma 12. □

### 5.1.1 Proof of Lemma 10

Note that, as per the update (4) of scheduling algorithm 1, the $\mathbf{r}(j)$ is such that

$$|\mathbf{r}(j)|_\infty \leq \sum_{k=1}^{j} \frac{1}{k} = O(\log j).$$

Therefore, the statement of Lemma 10 follows by establishing existence of $C$ so that for $j \geq C$,

$$\mathbb{E}\left[\|\widehat{\boldsymbol{\lambda}}(j) - \boldsymbol{\lambda}\|_1 + \|\widehat{\mathbf{s}}(j) - \mathbf{s}(\mathbf{r}(j))\|_1 | \mathbf{r}(j)\right] = O\left(\frac{1}{j^2}\right), \tag{52}$$

for $|\mathbf{r}(j)|_\infty = O(\log j)$. In the remaining proof, for simplicity of notation we will drop reference $\mathbf{r}(j)$ and simply use $\mathbb{E}[\cdot]$ in place of $\mathbb{E}[\cdot | \mathbf{r}(j)]$. We will establish that by arguing separately that $\mathbb{E}\left[\|\widehat{\boldsymbol{\lambda}}(j) - \boldsymbol{\lambda}\|_1\right] = O(1/j^2)$ and $\mathbb{E}[\|\widehat{\mathbf{s}}(j) - \mathbf{s}(\mathbf{r}(j))\|_1] = O(1/j^2)$.

First, we consider the deviation in $\widehat{\boldsymbol{\lambda}}(j)$. This will immediately follow from the property of arrival process. By definition $\widehat{\boldsymbol{\lambda}}(j)$ is the empirical arrival rate vector over $[L(j), L(j+1))$. Now for any $i$,

$$\begin{aligned}\widehat{\lambda}_i(j) &= \frac{1}{T(j)} A_i(L(j), L(j+1)) \\ &= \frac{1}{T(j)}\left(\sum_{k=1}^{T(j)} A_i(L(j) + k - 1, L(j) + k)\right).\end{aligned} \tag{53}$$

Now, $X_k \triangleq A_i(L(j) + k - 1, L(j) + k)$ are i.i.d. random variables with $\mathbb{E}[X_k] = \lambda_i$, bounded support $[0, K]$ and hence standard deviation at most $K$. Using this, we have

$$\begin{aligned}\mathbb{E}\left[|\widehat{\lambda}_i(j) - \lambda_i|\right] &= \left(\mathbb{E}\left[\frac{1}{T(j)}\left|\sum_{k=1}^{T(j)}(X_k - \mathbb{E}[X_k])\right|\right]\right) \\ &\leq \left(\mathbb{E}\left[\frac{1}{T^2(j)}\left(\sum_{k=1}^{T(j)}(X_k - \mathbb{E}[X_k])\right)^2\right]\right)^{1/2} \\ &\leq \left(\frac{1}{T(j)}\mathbb{E}[X_1^2]\right)^{1/2} \\ &\leq \frac{K}{\sqrt{T(j)}} \\ &= O(1/j^2),\end{aligned} \tag{54}$$

where the last inequality follows from $T(j) = \exp(\sqrt{j})$. This completes the proof of bound on deviation for $\widehat{\boldsymbol{\lambda}}(j)$.

Now, we consider deviations in $\widehat{\mathbf{s}}(j)$ compared to $\mathbf{s}(\mathbf{r}(j))$. For this, first we establish $\mathbb{E}[\widehat{\mathbf{s}}(j)]$ being close to $\mathbf{s}(\mathbf{r}(j))$ and then we establish $\widehat{\mathbf{s}}(j)$ being close to $\mathbb{E}[\widehat{\mathbf{s}}(j)]$. Therefore, we start by evaluating deviation between $\mathbb{E}[\widehat{\mathbf{s}}(j)]$ and $\mathbf{s}(\mathbf{r}(j))$. To this end, consider any $i$. We will establish that,

$$|\mathbb{E}[\widehat{s}_i(j)] - s_i(\mathbf{r}(j))| = O\left(\frac{1}{j^4}\right). \tag{55}$$

To establish (55), we will use the mixing time bounds (23) derived in Section 4.2 next. To this end, let $\boldsymbol{\mu}(t)$ be the distribution over $\mathcal{I}(G)$ of scheduling decisions at time $t \in [L(j), L(j+1))$. By Lemma 8(2), $s_i(\mathbf{r}(j)) = \mathbb{E}_{\boldsymbol{\pi}^{\mathbf{r}(j)}}[\sigma_i]$. And $\sigma_i$ is $0-1$ valued random variable. Therefore,

$$
\begin{aligned}
\left|\mathbb{E}_{\boldsymbol{\mu}(t)}[\sigma_i] - s_i(\mathbf{r}(j))\right| &= \left|\mathbb{E}_{\boldsymbol{\mu}(t)}[\sigma_i] - \mathbb{E}_{\boldsymbol{\pi}^{\mathbf{r}(j)}}[\sigma_i]\right| \\
&\leq \left|\boldsymbol{\mu}(t) - \boldsymbol{\pi}^{\mathbf{r}(j)}\right|_{TV} \\
&\leq \left\|\frac{\boldsymbol{\mu}(t)}{\boldsymbol{\pi}^{\mathbf{r}(j)}} - 1\right\|_{2, \boldsymbol{\pi}^{\mathbf{r}(j)}},
\end{aligned}
\quad (56)
$$

where the last inequality follows from (13). Now, from (23), the RHS of (56) is bounded above by $O(1/j^4)$ as long as

$$
\begin{aligned}
t &\geq L(j) + \left(\exp(\Theta(n|r|_{\max} + n))\log j^4\right) \\
&= L(j) + j^{O(n)}\log j = L(j) + \Gamma(j),
\end{aligned}
\quad (57)
$$

where $\Gamma(j) = j^{O(n)}\log j$. In above, while applying (23), we have used the fact $|\mathbf{r}|_\infty = O(\log j)$. This leads to the following bound.

$$
\begin{aligned}
|\mathbb{E}[\widehat{s}_i(j)] - s_i(\mathbf{r}(j))| &= \frac{1}{T(j)}\left|\int_{L(j)}^{L(j+1)} \mathbb{E}_{\boldsymbol{\mu}(t)}[\sigma_i] - \mathbb{E}_{\boldsymbol{\pi}^{\mathbf{r}(j)}}[\sigma_i]\, dt\right| \\
&\leq \frac{\Gamma(j)}{T(j)} + O\left(\frac{1}{j^4}\right) \\
&= O\left(\frac{1}{j^4}\right).
\end{aligned}
\quad (58)
$$

Hence the (55) follows since $(j^{O(n)}\log j)/T(j) = O(1/j^4)$ due to choice of $T(j) = \exp(\sqrt{j})$.

Given (55), as the last step to establish $\mathbb{E}[\|\widehat{\mathbf{s}}(j) - \mathbf{s}(\mathbf{r}(j))\|_1] = O(1/j^2)$, we will show that for any $i$,

$$
\mathbb{E}\left[|\widehat{s}_i(j) - \mathbb{E}[\widehat{s}_i(j)]|\right] = O\left(\frac{1}{j^2}\right).
\quad (59)
$$

Consider (with notation $S = [L(j), L(j+1))$),

$$
\begin{aligned}
T(j)^2 \mathbb{E}\left[|\widehat{s}_i(j) - \mathbb{E}[\widehat{s}_i(j)]|\right]^2 &= \mathbb{E}\left[\left|\int_{L(j)}^{L(j+1)} \sigma_i(t)\, dt - \mathbb{E}\left[\int_{L(j)}^{L(j+1)} \sigma_i(t)\, dt\right]\right|\right]^2 \\
&\leq \mathbb{E}\left[\left(\int_{L(j)}^{L(j+1)} \sigma_i(t)\, dt - \mathbb{E}\left[\int_{L(j)}^{L(j+1)} \sigma_i(t)\, dt\right]\right)^2\right] \\
&= \left(\mathbb{E}\left[\left(\int_{L(j)}^{L(j+1)} \sigma_i(t)\, dt\right)^2\right] - \mathbb{E}\left[\int_{L(j)}^{L(j+1)} \sigma_i(t)\, dt\right]^2\right) \\
&= \left(\int_{L(j)}^{L(j+1)} \mathbb{E}[\sigma_i(t)]\left(\int_{L(j)}^{L(j+1)} \mathbb{E}[\sigma_i(s)|\sigma_i(t)=1] - \mathbb{E}[\sigma_i(s)]\, ds\right) dt\right)
\end{aligned}
$$

$$
\begin{aligned}
&= 2\left(\int_{L(j)}^{L(j+1)} \mathbb{E}[\sigma_i(t)] \left(\int_t^{L(j+1)} \mathbb{E}[\sigma_i(s)|\sigma_i(t)=1] - \mathbb{E}[\sigma_i(s)]\ ds\right) dt\right) \\
&= 2\left(\int_{L(j)}^{L(j+1)} \mathbb{E}[\sigma_i(t)] \left(\int_t^{t+\Gamma(j)} \mathbb{E}[\sigma_i(s)|\sigma_i(t)=1] - \mathbb{E}[\sigma_i(s)]\ ds\right.\right. \\
&\qquad\qquad\left.\left. + \int_{t+\Gamma(j)}^{L(j+1)} \mathbb{E}[\sigma_i(s)|\sigma_i(t)=1] - \mathbb{E}[\sigma_i(s)]\ ds\right) dt\right) \\
&\stackrel{(a)}{\leq} 2\left(\int_{L(j)}^{L(j+1)} \mathbb{E}[\sigma_i(t)] \left(\int_t^{t+\Gamma(j)} 1\ ds + \int_{t+\Gamma(j)}^{L(j+1)} O\left(\frac{1}{j^4}\right) ds\right) dt\right) \\
&\leq 2T(j)\left(\Gamma(j) + O\left(\frac{T(j)}{j^4}\right)\right) \\
&= O\left(\frac{T(j)^2}{j^4}\right). \quad\quad\quad\quad\quad\quad\quad\quad\quad\quad\quad\quad\quad\quad\quad\quad (60)
\end{aligned}
$$

In above, (a) follows from choice of $\Gamma(j)$ as in (56), (57), if $s \geq t+\Gamma(j)$ then due to the 'mixing effect' $\mathbb{E}[\sigma_i(s)|\sigma_i(t)], \mathbb{E}[\sigma_i(s)]$ are within $O(1/j^4)$ of $s_i(\mathbf{r}(j))$. Now, (60) immediately implies that

$$\mathbb{E}\left[|\widehat{s}_i(j) - \mathbb{E}[\widehat{s}_i(j)]|\right] = O\left(\frac{1}{j^2}\right). \quad\quad (61)$$

To conclude, observe that (54), (55) and (61) imply the result of Lemma 10.

## 5.2 Proof of Theorem 2: Positive Harris Recurrence

The goal of this section is to prove Theorem 2, that is, the positive Harris recurrence of the network Markov process under Scheduling Algorithm 2. For a countable Markov chain, positive recurrence means that all states are visited infinitely often, with a finite mean inter-visit time. When the state space is not countable (as in our case), one cannot expect every state to be visited infinitely often. However, a small set of states can have that property. If the transition probabilities out of that set are similar, then the set plays the role of a recurrent state. Indeed, the evolution essentially starts afresh once the chain hits that set. This idea is made precise by the definition of a petite[2] set. Section 4.3 has review of known results about establishing positive Harris recurrence. In particular, Theorem 6 there states that the existence of a positive recurrent closed petite set implies positive Harris recurrence.

The appropriate petite set is the set $\mathbf{S}$ where the sum of the squares of the queue lengths is less than some constant $\kappa$. The positive recurrence is proved using the fact that the sum of the squares of the queue lengths is a Lyapunov function which tends to decrease when it is larger than $\kappa$ (Lemma 13). Intuitively, this is true because Scheduling Algorithm 2 tries to balance $\widehat{s}_i(j)$ and $\widehat{\lambda}_i(j)+\varepsilon$ for all $i$, so that on average, the service rate dominates the arrival rate on each queue. The set $\mathbf{S}$ is shown to be petite (Lemma 14) by proving that starting from any state in that set, there is some lower bound $\theta$ on the probability that, at some later time $T_\kappa$, the queues become empty, no link is active, and the parameters **r** of the CSMA backoff delays reach their maximum value (Proposition 17). Thus, the evolution of the Markov chain essentially starts afresh from that set with at least probability $\theta$.

---
[2] Recall that petite means small in French.

To this end, we start with necessary definitions of the network Markov process under scheduling algorithm 2. Let $\tau \in \mathbb{N} \cup \{0\}$ be the index for the discrete time. It can be checked that the tuple $X(j) = (\mathbf{Q}(Tj), \mathbf{r}(Tj), \sigma(Tj))$ forms the state of the time-homogeneous Markov chain operating under the algorithm. Now $X(\tau) \in \mathsf{X}$ where $\mathsf{X} = \mathbb{R}_+^n \times [-\frac{n}{\varepsilon}, \frac{n}{\varepsilon}]^n \times \mathcal{I}(G)$. Clearly, $\mathsf{X}$ is a Polish space endowed with the natural product topology. Let $\mathcal{B}_\mathsf{X}$ be the Borel $\sigma$-algebra of $\mathsf{X}$ with respect to this product topology. Finally, for $\mathbf{x} = (\mathbf{Q}, \mathbf{r}, \sigma) \in \mathsf{X}$, define norm of $\mathbf{x}$ denoted by $|\mathbf{x}|$ as

$$|\mathbf{x}| = |\mathbf{Q}| + |\mathbf{r}| + |\sigma|,$$

where $|\mathbf{Q}|, |\mathbf{r}|$ and $|S|$ denotes the $\ell_1$ norm, $|\sigma|$ is its index in $\{0, \ldots, |\mathcal{I}(G)| - 1\}$, assigned arbitrarily. Thus, $|\mathbf{r}|, |\sigma|$ are always bounded. Therefore, in essence $|\mathbf{x}| \to \infty$ iff $|\mathbf{Q}| \to \infty$.

To establish statement of Theorem 2, we need to show that $X(\tau)$ is indeed positive Harris recurrent as long as $\boldsymbol{\lambda} + 2\varepsilon\mathbf{1} \in \boldsymbol{\Lambda}$. By Theorem 6, it is sufficient to find positive recurrent closed petit set. First, we will find closed recurrent set using criterion of Lemma 7 and then establish that the set is indeed petit. To this end, define a Lyapunov function $L : \mathsf{X} \to \mathbb{R}_+$ as

$$L(\mathbf{x}) = \sum_{i=1}^n Q_i^2 \triangleq \mathbf{Q}^2 \cdot \mathbf{1}, \quad \text{where} \quad \mathbf{x} = (\mathbf{Q}, \mathbf{r}, \boldsymbol{\sigma}) \in \mathsf{X}.$$

We establish the following 'drift' property about $L$.

**Lemma 13** *Given $\boldsymbol{\lambda}$ so that $\boldsymbol{\lambda} + 2\varepsilon\mathbf{1} \in \boldsymbol{\Lambda}$, define*

$$N = N(\varepsilon, n) = \left\lceil \frac{48 \times 16 \times 72 n^5}{\varepsilon^6} \right\rceil.$$

*Then, for any initial state $X(0) = (\mathbf{Q}(0), \mathbf{r}(0), \boldsymbol{\sigma}(0)) \in \mathsf{X}$,*

$$\mathbb{E}\left[L(X(N)) - L(X(0)) | X(0)\right] \leq -h(X(0)), \tag{62}$$

*where $h : \mathsf{X} \to \mathbb{R}$ is defined as*

$$h(\mathbf{x}) = \varepsilon TN(\mathbf{Q}(0) \cdot \mathbf{1}) - n(TN)^2 \left(\varepsilon + K^2 + 2K\right). \tag{63}$$

Therefore, Lemma 7 implies that for some finite $\kappa > 0$, set $B_\kappa = \{\mathbf{x} : L(\mathbf{x}) \leq \kappa\}$ satisfies

$$\mathbb{E}_\mathbf{x}[T_{B_\kappa}] < \infty, \quad \text{for any } \mathbf{x} \in \mathsf{X}$$
$$\sup_{\mathbf{x} \in B_\kappa} \mathbb{E}_\mathbf{x}[T_{B_\kappa}] < \infty.$$

Therefore, by Theorem 6 the following is sufficient to complete the proof of Theorem 2.

**Lemma 14** *Consider any $\kappa > 0$. Then, the set $B_\kappa = \{\mathbf{x} : L(\mathbf{x}) \leq \kappa\}$ is a closed petite set.*

In the remainder of this sub-section, we shall prove Lemmas 13 and 14.

### 5.2.1 Proof of Lemma 13

**A relevant optimization problem.** The basic idea behind the update algorithm (6) is to design a simple gradient procedure for solving the following optimization problem.

$$\begin{aligned} \text{maximize} \quad & F(\mathbf{r}, \boldsymbol{\lambda} + \varepsilon \mathbf{1}) \triangleq F_\varepsilon(\mathbf{r}) \\ \text{subject to} \quad & \mathbf{r} \in \mathbb{R}^n. \end{aligned} \quad (64)$$

By Lemma 8, it follows that if $\boldsymbol{\lambda} + 2\varepsilon \mathbf{1} \in \Lambda$, then (64) has a unique solution that is attained; let it be $\mathbf{r}^* = \mathbf{r}^*(\boldsymbol{\lambda} + \varepsilon \mathbf{1})$. Then, from Lemma 8(2) the effective service rate $\mathbf{s}(\mathbf{r}^*)$, under the random access algorithm with fixed $\mathbf{r}^*$, is such that

$$s_i(\mathbf{r}^*) = \lambda_i + \varepsilon.$$

That is, the arrival rate is less than the service rate by $\varepsilon > 0$ under this idealized setup. In order to establish the positive Harris recurrence, we will need more than this – service rate should dominate arrival rate for small enough time interval to imply appropriate drift condition desired by Lyapunov-Foster's criteria. This is exactly what we will establish next.

**Derivative of $F_\varepsilon$ becomes small.** As per statement of Lemma 13, let initial state be $X(0) = (\mathbf{Q}(0), \mathbf{r}(0), \boldsymbol{\sigma}(0))$. As the first step, we wish to establish the following:

$$\begin{aligned} \frac{1}{N} \sum_{j=1}^N \mathbb{E}\left[\|\nabla F_\varepsilon(\mathbf{r}(j))\|_2^2\right] &= \frac{1}{N} \sum_{j=1}^N \mathbb{E}\left[\|\boldsymbol{\lambda} + \varepsilon \cdot \mathbf{1} - \mathbf{s}(\mathbf{r}(j))\|_2^2\right] \\ &\leq \frac{\varepsilon^2}{16}. \end{aligned} \quad (65)$$

In the above and everywhere else in the proof of Lemma 13, the expectation is always assumed to be conditioned on the initial state $X(0)$. For simplicity we will drop reference to this conditioning. Intuitively, (65) implies that on average and in expectation, the arriving rate $\boldsymbol{\lambda}$ is strictly less than the normalized service rate $\mathbf{s}(\mathbf{r}(j))$ after a finite time $N$. This will allow us to establish drift in Lyapunov function. To this end, we start with definition $G(\mathbf{r}) = F_\varepsilon(\mathbf{r}) - \|\mathbf{r} - \mathbf{r}^*\|_2^2$. We establish the follwing useful non-decreasing property of $G(\cdot)$ under the 'projection' defined in (7).

**Lemma 15** *For any $\mathbf{r} \in \left[-\frac{n}{\varepsilon}, \frac{n}{\varepsilon}\right]^n$ and $\Delta \mathbf{r} \in [-1,1]^n$, $-\frac{16n^3}{\varepsilon^2} \leq G(\mathbf{r}) < 0$ and $G([\mathbf{r} + \Delta \mathbf{r}]_{\frac{n}{\varepsilon}}) \geq G(\mathbf{r} + \Delta \mathbf{r})$.*

*Proof.* $G(\mathbf{r})$ is upper bounded by 0 since $F_\varepsilon(\mathbf{r}) \leq F_\varepsilon(\mathbf{r}^*) < 0$ by Lemma 8. Further,

$$\begin{aligned} G(\mathbf{r}) &= F_\varepsilon(\mathbf{r}) - \|\mathbf{r} - \mathbf{r}^*\|_2^2 \\ &\stackrel{(a)}{\geq} (\boldsymbol{\lambda} + \varepsilon \cdot \mathbf{1}) \cdot \mathbf{r} - \log\left(\sum_{\sigma \in \mathcal{I}(G)} \exp(\sigma \cdot \mathbf{r})\right) - n\left(\frac{2n}{\varepsilon}\right)^2 \\ &\geq n \cdot \left(-\frac{n}{\varepsilon}\right) - \log\left(2^n \exp(n\mathbf{r}_{\max})\right) - n\left(\frac{2n}{\varepsilon}\right)^2 \\ &\geq -\frac{16n^3}{\varepsilon^2}. \end{aligned} \quad (66)$$

Here (a) follows from Lemma 8(3) for $\mathbf{r}^* = \mathbf{r}^*(\boldsymbol{\lambda} + \varepsilon \mathbf{1})$ (thus $\mathbf{r}^* \in [-\frac{n}{\varepsilon}, \frac{n}{\varepsilon}]^n$), and the last step has used $n > 3$. Now if we set $\mathbf{x} = \mathbf{r} + \Delta \mathbf{r}$, $|\mathbf{x}|_{\max} \leq \frac{n}{\varepsilon} + 1$ and we need to show $G([\mathbf{x}]_{\frac{n}{\varepsilon}}) \geq G(\mathbf{x})$. Note that it is enough to show that for any $i \in V$,

$$G([\mathbf{x}]_{\frac{n}{\varepsilon}, i}) \geq G(\mathbf{x}), \tag{67}$$

where the $i$-projection $\bar{\mathbf{x}} = [\mathbf{x}]_{\frac{n}{\varepsilon}, i}$ is defined as

$$\bar{x}_j = \begin{cases} \left([\mathbf{x}]_{\frac{n}{\varepsilon}}\right)_j & \text{if } j = i \\ x_j & \text{otherwise} \end{cases}.$$

Then we can iteratively apply (67) to complete the proof. When $x_i \in [-\frac{n}{\varepsilon}, \frac{n}{\varepsilon}]$, desired claim trivially follows as $[\mathbf{x}]_{\frac{n}{\varepsilon}, i} = \mathbf{x}$. Now suppose $x_i \notin [-\frac{n}{\varepsilon}, \frac{n}{\varepsilon}]$. By definition, it must be that $x_i \in (\frac{n}{\varepsilon}, \frac{n}{\varepsilon} + 1]$ or $x_i \in [-\frac{n}{\varepsilon} - 1, -\frac{n}{\varepsilon})$. We prove (67) when $x_i \in [\frac{n}{\varepsilon}, \frac{n}{\varepsilon} + 1]$; the other arguments for the other case are very similar. Consider,

$$
\begin{aligned}
G([\mathbf{x}]_{\frac{n}{\varepsilon}, i}) - G(\mathbf{x}) &= F_\varepsilon([\mathbf{x}]_{\frac{n}{\varepsilon}, i}) - F_\varepsilon(\mathbf{x}) - \left(\frac{n}{\varepsilon} - r_i^*\right)^2 + (x_i - r_i^*)^2 \\
&\overset{(a)}{\geq} -\left(x_i - \frac{n}{\varepsilon}\right) + \left(x_i - \frac{n}{\varepsilon}\right)\left(x_i + \frac{n}{\varepsilon} - 2r_i^*\right) \\
&\overset{(b)}{\geq} -\left(x_i - \frac{n}{\varepsilon}\right) + 2\left(x_i - \frac{n}{\varepsilon}\right) \\
&\geq 0.
\end{aligned}
$$

In above, (a) and (b) is due to $\left|\frac{\partial F_\varepsilon}{\partial r_i}\right| \leq 1$ and $|\mathbf{r}^*|_{\max} \leq \frac{n \cdot \log(2)}{\varepsilon} < \frac{n}{\varepsilon} - 1$ (since $n > 3$ by assumption), respectively. This completes the proof of Lemma 15. $\square$

Now consider the relation between $G(\mathbf{r}(j+1))$ and $G(\mathbf{r}(j))$.

$$
\begin{aligned}
G(\mathbf{r}(j+1)) &= G\left([\mathbf{r}(j) + \alpha \left(\nabla F_\varepsilon(\mathbf{r}(j)) + e(j)\right)]_{\frac{n}{\varepsilon}}\right) \\
&\overset{(a)}{\geq} G\left(\mathbf{r}(j) + \alpha \left(\nabla F_\varepsilon(\mathbf{r}(j)) + e(j)\right)\right) \\
&= F_\varepsilon\left(\mathbf{r}(j) + \alpha \left(\nabla F_\varepsilon(\mathbf{r}(j)) + e(j)\right)\right) - \|\mathbf{r}(j) + \alpha \left(\nabla F_\varepsilon(\mathbf{r}(j)) + e(j)\right) - \mathbf{r}^*\|_2^2 \\
&= F_\varepsilon(\mathbf{r}(j)) + \nabla F_\varepsilon(\mathbf{r}(j)) \cdot \alpha \left(\nabla F_\varepsilon(\mathbf{r}(j)) + e(j)\right) \\
&\quad + \frac{1}{2}\alpha \left(\nabla F_\varepsilon(\mathbf{r}(j)) + e(j)\right) \cdot M \cdot \alpha \left(\nabla F_\varepsilon(\mathbf{r}(j)) + e(j)\right) \\
&\quad - \|\mathbf{r}(j) - \mathbf{r}^*\|_2^2 - 2\alpha \nabla F_\varepsilon(\mathbf{r}(j)) \cdot (\mathbf{r}(j) - \mathbf{r}^*) - 2\alpha \cdot e(j) \cdot (\mathbf{r}(j) - \mathbf{r}^*) \\
&\quad - \alpha^2 \|\nabla F_\varepsilon(\mathbf{r}(j)) + e(j)\|_2^2 \\
&\overset{(b)}{\geq} F_\varepsilon(\mathbf{r}(j)) + \alpha \|\nabla F_\varepsilon(\mathbf{r}(j))\|_2^2 + \alpha \nabla F_\varepsilon(\mathbf{r}(j)) \cdot e(j) - \frac{\alpha^2 (K+1)^2 n^2}{2} \\
&\quad - \|\mathbf{r}(j) - r^*\|_2^2 - 2\alpha \cdot e(j) \cdot (\mathbf{r}(j) - \mathbf{r}^*) - \alpha^2 (K+1)^2 n^2 \\
&\overset{(c)}{\geq} F_\varepsilon(\mathbf{r}(j)) + \alpha \|\nabla F_\varepsilon(\mathbf{r}(j))\|_2^2 - \alpha \|e(j)\|_1 - \frac{3\alpha^2 (K+1)^2 n^2}{2} \\
&\quad - \|\mathbf{r}(j) - \mathbf{r}^*\|_2^2 - 2\alpha \|e(j)\|_1 \times \frac{2n}{\varepsilon}
\end{aligned}
$$

$$\begin{aligned}
&= G(\mathbf{r}(j)) + \alpha\|\nabla F_\varepsilon(\mathbf{r}(j))\|_2^2 - \alpha\left(1 + \frac{4n}{\varepsilon}\right)\|e(j)\|_1 - \frac{3\alpha^2(K+1)^2 n^2}{2} \\
&\geq G(\mathbf{r}(j)) + \alpha\|\nabla F_\varepsilon(\mathbf{r}(j))\|_2^2 - \frac{5\alpha n}{\varepsilon}\|e(j)\|_1 - \frac{3\alpha^2(K+1)^2 n^2}{2}, \quad (68)
\end{aligned}$$

where the random vector $e(j) = \widehat{\lambda}_i(j) - \widehat{s}_i(j) - (\lambda_i - s_i(\mathbf{r}(j)))$; $M$ is the $n \times n$ with $M_{ab} = \partial^2 F_\varepsilon(\tilde{\mathbf{r}})/\partial r_a \partial r_b$ for some $\tilde{\mathbf{r}}$ in neighborhood of $\mathbf{r}$ with $M_{ab} \in [-1,1]$ by (40). In above (a) follows from the fact that $\alpha \leq (K+1)^{-2}$, $\nabla F_\varepsilon(\mathbf{r}(j)) \in [-1,1]^n$, $e(j) \in [-1,K]^n$ and Lemma 15[3]. For (b), we use that $|M|_\infty \leq 1$ and the concavity of $F_\varepsilon$ and $\nabla F_\varepsilon(\mathbf{r}(j)) + e(j) \in [-2,(K+1)]^n$. Finally (c) follows from $\nabla F_\varepsilon(\mathbf{r}(j)) \in [-1,1]^n$ and $\mathbf{r}(j) - \mathbf{r}^* \in \left[-\frac{2n}{\varepsilon}, \frac{2n}{\varepsilon}\right]^n$.

Our choice of the large updating period $T$ is merely for bounding $e(j)$ and we obtain the following lemma which is analogous to Lemma 10.

**Lemma 16** *If the updating period $T \geq \exp\left(\Theta\left(\frac{n^2}{\varepsilon}\log\frac{n}{\varepsilon}\right)\right)$, then for all $j \in \mathbb{N}$*

$$\mathbb{E}\left[\|\widehat{\boldsymbol{\lambda}}(j) - \boldsymbol{\lambda}\|_1 + \|\widehat{\mathbf{s}}(j) - \mathbf{s}(\mathbf{r}(j))\|_1\right] \leq \frac{\varepsilon^3}{240n}.$$

*Therefore, for all $j \in \mathbb{N}$*

$$\mathbb{E}\left[\|e(j)\|_1\right] \leq \frac{\varepsilon^3}{240n}.$$

*Proof.* We provide sketch proof here since the proof of Lemma 16 is essentially the same as that of Lemma 10 – replace $T(j) = T$, $\alpha(j) = \alpha$ for all $j$ and use $|\mathbf{r}|_{\max} \leq \frac{n}{\varepsilon}$ to obtain bound of

$$\exp\left(\Theta\left(\frac{n^2}{\varepsilon}\right)\right),$$

on mixing time of the Markov chain on $\mathcal{I}(G)$ using (5). As a consequence, it follows that by choice of $T$ with large enough constant in its exponent, as stated in Lemma 16, the expectation of $\|e(j)\|_1$ can be made smaller than any given constant. Specifically, it can be made smaller than $\frac{\varepsilon^3}{240n}$. □

Summing (68) from $j = 1$ to $N$,

$$\begin{aligned}
0 &\geq G(\mathbf{r}(N+1)) \\
&\geq G(\mathbf{r}(1)) + \alpha\left(\sum_{j=1}^N \|\nabla F_\varepsilon(\mathbf{r}(j))\|_2^2\right) - \frac{5\alpha n}{\varepsilon}\left(\sum_{j=1}^N \|e(j)\|_1\right) - \frac{3\alpha^2(K+1)^2 n^2}{2}N. \quad (69)
\end{aligned}$$

Taking expectation on both sides and diving by $\alpha N$,

$$\begin{aligned}
\frac{1}{N}\sum_{j=1}^N E\left[\|\nabla F_\varepsilon(\mathbf{r}(j))\|_2^2\right] &\leq -\frac{1}{\alpha N}G(\mathbf{r}(1)) + \frac{5n}{\varepsilon N}\sum_{j=1}^N E[\|e(j)\|_1] + \frac{3\alpha(K+1)^2 n^2}{2} \\
&\stackrel{(a)}{\leq} \frac{1}{\alpha N}\frac{16n^3}{\varepsilon^2} + \frac{\varepsilon^2}{48} + \frac{\varepsilon^2}{48} \\
&\leq \frac{\varepsilon^2}{16}, \quad (70)
\end{aligned}$$

---
[3] This is the main reason why we consider $G$ instead of $F_\varepsilon$ as we can not establish monotonicity of $F_\varepsilon$ under the projection.

since $N = \left\lceil \frac{48 \times 16 n^3}{\alpha \varepsilon^4} \right\rceil = \left\lceil \frac{48 \times 16 \times 72 n^5}{\varepsilon^6} \right\rceil$, $\alpha = \frac{\varepsilon^2}{72(K+1)^2 n^2}$ and Lemmas 15 and 16.

**Service rate dominates arrival rate.** Next, we wish to establish that the average of empirical service rate dominates the average arrival rate over time interval of length $N$. That is, for all $i$

$$\frac{1}{N}\left(\sum_{j=1}^{N}(\mathbb{E}[\widehat{s}_i(j)])\right) \geq \lambda_i + \varepsilon/2. \tag{71}$$

To this end, first note that

$$\frac{1}{N}\sum_{j=1}^{N}\mathbb{E}\left[\left|\frac{\partial F_\varepsilon}{\partial r_i}(\mathbf{r}(j))\right|\right] \leq \sqrt{\frac{1}{N}\sum_{j=1}^{N}\mathbb{E}\left[\left|\frac{\partial F_\varepsilon}{\partial r_i}(\mathbf{r}(j))\right|\right]^2}, \quad \text{(from Cauchy-Schwarz inequality)}$$

$$\leq \sqrt{\frac{1}{N}\sum_{j=1}^{N}\mathbb{E}\left[\left|\frac{\partial F_\varepsilon}{\partial r_i}(\mathbf{r}(j))\right|^2\right]} \leq \frac{\varepsilon}{4}, \tag{72}$$

where the last inequality is from (70). Therefore,

$$\begin{aligned}
\frac{1}{N}\left(\sum_{j=1}^{N}(\mathbb{E}[\widehat{s}_i(j)] - \lambda_i)\right) &= \frac{1}{N}\left(\sum_{j=1}^{N}((\mathbb{E}[s_i(\mathbf{r}(j))] - \lambda_i) + (\mathbb{E}[\widehat{s}_i(j)] - \mathbb{E}[s_i(\mathbf{r}(j))]))\right) \\
&\geq \frac{1}{N}\left(\sum_{j=1}^{N}((\mathbb{E}[s_i(\mathbf{r}(j))] - \lambda_i) - |\mathbb{E}[\widehat{s}_i(j)] - \mathbb{E}[s_i(\mathbf{r}(j))]|)\right) \\
&\stackrel{(a)}{\geq} \frac{1}{N}\left(\sum_{j=1}^{N}\left(\varepsilon - \mathbb{E}\left[\frac{\partial F_\varepsilon}{\partial r_i}(\mathbf{r}(j))\right] - \frac{\varepsilon^3}{240n}\right)\right) \\
&\geq \frac{3}{4}\varepsilon - \frac{1}{N}\sum_{j=1}^{N}\mathbb{E}\left[\left|\frac{\partial F_\varepsilon}{\partial r_i}(\mathbf{r}(j))\right|\right] \\
&\stackrel{(b)}{\geq} \frac{\varepsilon}{2}, \tag{73}
\end{aligned}$$

In above, (a) follows from Lemma 8(2), i.e.

$$s_i(\mathbf{r}(j)) = \lambda_i + \varepsilon - \frac{\partial F_\varepsilon}{\partial r_i}(\mathbf{r}(j)),$$

and from Lemma 16. The (b) follows from (72).

**Wrapping up: Negative drift.** Now, consider $Q_i(N)$. For this, suppose $Q_i(0) > TN$. Then, $Q_i(\cdot)$ is strictly positive over interval $[0, TN]$ as service rate is at most 1. Therefore, in that case the queue $Q_i(\cdot)$ is *fully* served in time $[0, TN]$. Hence, using (73), we conclude that

$$\begin{aligned}
\mathbb{E}[Q_i(TN)] &= Q_i(0) + T\left(\sum_{j=1}^{N}\mathbb{E}[\lambda_i - \widehat{s}_i(j)]\right) \\
&\leq Q_i(0) - \frac{\varepsilon}{2}T \cdot N, \tag{74}
\end{aligned}$$

if $Q_i(0) > TN$. In above, as usual we have assumed that the expectation is conditional with respect to $X(0)$. In what follows, we will use this conditioning explicitly. Given (74), we have

$$\begin{aligned}
\mathbb{E}[Q_i^2(TN) - Q_i^2(0)|X(0)] &= \mathbb{E}[(Q_i(TN) - Q_i(0))(Q_i(TN) + Q_i(0))|X(0)] \\
&= \mathbb{E}[(Q_i(TN) - Q_i(0))^2 + 2Q_i(0)(Q_i(TN) - Q_i(0))|X(0)] \\
&\stackrel{(a)}{\leq} (KTN)^2 + 2Q_i(0)\mathbb{E}[Q_i(TN) - Q_i(0)|X(0)] \\
&\stackrel{(b)}{\leq} \begin{cases} (KTN)^2 - \frac{\varepsilon}{2}TN \times 2Q_i(0) & \text{if } Q_i(0) > TN \\ (K^2 + 2K)(TN)^2 & \text{if } Q_i(0) \leq TN \end{cases} \quad (75) \\
&\leq -\varepsilon TN Q_i(0) + \varepsilon(TN)^2 + (K^2 + 2K)(TN)^2, \quad (76)
\end{aligned}$$

for all $\mathbf{Q}(0)$. In above, (a) is from boundedness of arrival process and (b) is from (74). Hence,

$$\begin{aligned}
\mathbb{E}[L(X(N)) - L(X(0))|X(0)] &= \mathbb{E}\left[\sum_{i=1}^n Q_i^2(TN) - \sum_{i=1}^n Q_i^2(0) \Big| X(0)\right] \\
&\leq -\varepsilon TN \left(\sum_{i=1}^n Q_i(0)\right) + \varepsilon n (TN)^2 + n(K^2 + 2K)(TN)^2.
\end{aligned}$$

This completes the proof of Lemma 13.

## 5.3 Proof of Lemma 14

We wish to establish that set $B_\kappa = \{\mathbf{x} : L(\mathbf{x}) \leq \kappa\}$ is a closed petit set. By definition, it is closed. To establish that it is a petit set, we need to find a non-trivial measure $\mu$ on $(\mathsf{X}, \mathcal{B}_\mathsf{X})$ and a sampling distribution $a$ on $\mathbb{N}$ so that for any $\mathbf{x} \in B_\kappa$,

$$K_a(\mathbf{x}, \cdot) \geq \mu(\cdot).$$

To construct such a measure $\mu$, we shall use the following Proposition.

**Proposition 17** *Let the network Markov chain $X(\cdot)$ start with state $\mathbf{x} \in B_\kappa$ at time 0, $X(0) = \mathbf{x}$. Then, there exists $T_\kappa \geq 1$ and $\gamma_\kappa > 0$ such that*

$$\sum_{\tau=1}^{T_\kappa} \Pr_\mathbf{x}(X(\tau) = \mathbf{y}) \geq \gamma_\kappa, \quad \forall \mathbf{x} \in B_\kappa.$$

*Here $\mathbf{y} = (\mathbf{0}, [\frac{n}{\varepsilon}], \mathbf{0}) \in \mathsf{X}$ denote the state where all components of $\mathbf{Q}$ and $\sigma$ (i.e. the schedule is the empty independent set) and $r_i = \frac{n}{\varepsilon}$ for all $i \in V$.*

*Proof.* Consider any $\mathbf{x} \in B_\kappa$. By definition total amount of work in each queue is no more than $\sqrt{\kappa} + 1$. Consider some large enough (soon to be determined) $T_\kappa$. By the property of the assumed arrival process, there is a positive probability $\theta_\kappa^0 > 0$ of no arrivals happening to the system in time $T_\kappa$. Assuming no arrivals happen, we will show that in large enough time $t_\kappa^1$, with probability $\theta_\kappa^1 > 0$ each queue receives at least $\sqrt{\kappa} + 1$ amount of service; and after that in additional time $t^2$ with positive probability $\theta^2 > 0$ the empty set schedule is reached. Now, after the empty set schedule is reached, in additional time $t^3$ with positive probability $\theta^3 > 0$, the empty set schedule remains; i.e. the scheduling does not change in this time. Since the empty set schedule remains and no packet arrives, $r_i$ is increasing by $\varepsilon$ from (6) and finally reach $\frac{n}{\varepsilon}$ for a large enough $t^3$ which depends on $n$. This will imply that by defining $T_\kappa \triangleq t^1_\kappa + t^2 + t^3$ the state $\mathbf{y} \in \mathsf{X}$ is reached with probability at least

$$\gamma_\kappa \triangleq \theta^0_\kappa \theta^1_\kappa \theta^2 \theta^3 > 0.$$

And this will immediately imply the desired result of Proposition 17. To this end, we need to show existence of $t^1_\kappa, t^2, t^3$ and $\theta^1_\kappa, \theta^2, \theta^3$ with properties stated above to complete the proof of Proposition 17.

First, show the existence of $t^1_\kappa, \theta^1_\kappa$. For this, note that the Markov chain corresponding to the scheduling algorithm has always bounded transition probabilities (since $\mathbf{r}$ is bounded in terms of $n$) and is irreducible over the space of all independent sets $\mathcal{I}(G)$. Therefore, it follows that starting from any initial scheduling configuration, there exists finite time $\widehat{t}$ such that a schedule is reached so that any given queue $i$ is scheduled for at least unit amount of time with probability at least $\widehat{\theta} > 0$. Here, both $\widehat{t}, \widehat{\theta}$ depend on only $n$ (and $\varepsilon$), not $\kappa$. Therefore, it follows that in time $t^1_\kappa \triangleq (\sqrt{\kappa}+1)n\widehat{t}$ all queues become empty with probability at least $\theta^1_\kappa \triangleq \left(\widehat{\theta}\right)^{n(\sqrt{\kappa}+1)}$. Next, the existence of $t^2, \theta^2$ is also follows from the bounded property of our Markov chain. Finally, for $t^3, \theta^3$, consider the interpretation of the Markov chain as in Section 4.2 using the clock ticks. Note that no clock ticks in time $t^3$ with probability $\theta^3 > 0$ since its rate is bounded in terms of $n$. Hence, the empty set schedule remains in time $t^3$ with probability $\theta^3 > 0$, where $t^3$ and $\theta^3$ depends only on $n$. This completes the proof of Proposition 17. $\square$

In what follows, Proposition 17 will be used to complete the proof of Lemma 14. To this end, consider Geometric(1/2) as the sampling distribution $a$, i.e.

$$a(\ell) = 2^{-\ell}, \quad \ell \geq 1.$$

Let $\boldsymbol{\delta}_\mathbf{y}$ be the delta distribution on element $\mathbf{y} \in \mathsf{X}$. Then, define $\mu$ as

$$\mu = 2^{-T_\kappa} \gamma_k \boldsymbol{\delta}_\mathbf{y}, \quad \text{that is} \quad \mu(\cdot) = 2^{-T_\kappa} \gamma_k \boldsymbol{\delta}_\mathbf{y}(\cdot).$$

Clearly, $\mu$ is non-trivial measure on $(\mathsf{X}, \mathcal{B}_\mathsf{X})$. With these definitions of $a$ and $\mu$, Proposition 17 immediately implies that for any $\mathbf{x} \in B_\kappa$,

$$K_a(\mathbf{x}, \cdot) \geq \mu(\cdot).$$

This establishes that set $B_\kappa$ is a closed petit set and this completes the proof of Lemma 14.

# 6 Throughput & Fairness of Congestion Control Algorithms

## 6.1 Proof of Theorem 3: Rate Stable Congestion Control

The proof of Theorem 3 is similar to that of Theorem 1. In a nutshell, the basic idea is to show that the update equation (8) solves a relevant optimization problem through a subgradient algorithm. That is, $\boldsymbol{\lambda}(j), \mathbf{r}(j)$ converge to the solution of the appropriate optimization problem with probability 1. The property of the optimization problem will imply the goodness of utility of the convergent arrival rates. And, using this convergence property, it will in turn imply rate stability of queue-size.

**A relevant optimization problem & its properties.** Let $\mathcal{M}$ be space of all probability distributions on $\mathcal{I}(G)$. Given a distribution $\boldsymbol{\mu} \in \mathcal{M}$, by $\mathrm{H}_{ER}(\boldsymbol{\mu})$ denote its entropy defined as

$$\mathrm{H}_{ER}(\boldsymbol{\mu}) = - \sum_{\boldsymbol{\sigma} \in \mathcal{I}(G)} \mu_{\boldsymbol{\sigma}} \log \mu_{\boldsymbol{\sigma}}.$$

Consider the following optimization problem.

$$\begin{aligned} \text{maximize} \quad & \mathrm{H}_{ER}(\boldsymbol{\mu}) + \beta \left( \sum_i U_i(\lambda_i) \right) \quad \text{over} \quad \boldsymbol{\mu} \in \mathcal{M}, \ \boldsymbol{\lambda} \in [0,1]^n \\ \text{subject to} \quad & \mathbb{E}_{\boldsymbol{\mu}}[\sigma_i] \geq \lambda_i, \quad \text{for all } i. \end{aligned} \quad (77)$$

Associate a dual variable $r_i \geq 0$ to constraint $\mathbb{E}_{\boldsymbol{\mu}}[\sigma_i] \geq \lambda_i$. Here the use of $r_i$ for dual variable is an intentional abuse of notation and the reason behind this will soon become clear to the reader. Given this, the result Lagrangian is given by

$$\begin{aligned} \mathcal{L}(\boldsymbol{\mu}, \boldsymbol{\lambda}; \mathbf{r}) &= \mathrm{H}_{ER}(\boldsymbol{\mu}) + \beta \left( \sum_i U_i(\lambda_i) \right) + \left( \sum_i r_i (\mathbb{E}_{\boldsymbol{\mu}}[\sigma_i] - \lambda_i) \right) \\ &= \left( \mathrm{H}_{ER}(\boldsymbol{\mu}) + \sum_i r_i \mathbb{E}_{\boldsymbol{\mu}}[\sigma_i] \right) + \left( \sum_i [\beta U_i(\lambda_i) - r_i \lambda_i] \right). \end{aligned} \quad (78)$$

And, therefore the dual function is given by

$$\mathcal{D}(\mathbf{r}) = \sup \mathcal{L}(\boldsymbol{\mu}, \boldsymbol{\lambda}; \mathbf{r}) \quad \text{over} \quad \boldsymbol{\mu} \in \mathcal{M}, \ \boldsymbol{\lambda} \in [0,1]^n. \quad (79)$$

Finally, the dual optimization of (77) is given by

$$\text{minimize} \quad \mathcal{D}(\mathbf{r}) \quad \text{over } \mathbf{r} \in \mathbb{R}_+^n. \quad (80)$$

Now we are ready to state useful properties of the optimization problems, (77) and (80). These properties were present in earlier work [26].

**Lemma 18** *The optimization problem (77) is concave maximization while the optimization problem (80) is convex minization. There is no duality gap and hence both same the same optimal cost. They satisfy the following properties.*

(1) *Given dual feasible $\mathbf{r} \in \mathbb{R}_+^n$, the associate primal feasible assignment $\boldsymbol{\mu}(\mathbf{r}), \boldsymbol{\lambda}(\mathbf{r})$ are given as follows:*

$$\mu_{\boldsymbol{\sigma}} \ \propto \ \exp(\boldsymbol{\sigma} \cdot \mathbf{r}), \quad \text{for all} \ \boldsymbol{\sigma} \in \mathcal{I}(G). \quad (81)$$

*That is, $\boldsymbol{\mu}(\mathbf{r}) = \boldsymbol{\pi}^{\mathbf{r}}$. And*

$$\lambda_i(\mathbf{r}) = \arg\max_{y \in [0,1]} \left( \beta U_i(y) - r_i y \right), \quad \text{for all} \ i. \quad (82)$$

(2) The subgradient for $\mathcal{D}(\mathbf{r})$, represented as $g(\mathbf{r}) = [g_i(\mathbf{r})]$ is given by

$$g_i(\mathbf{r}) = \mathbb{E}_{\boldsymbol{\mu}(\mathbf{r})}[\sigma_i] - \lambda_i(\mathbf{r}).$$

(3) And, both problems have unique optimal solutions.

*Proof.* To begin with observe that the objective of (77) is strictly concave as entropy is a strictly concave function over $\mathcal{M}$ and so are $U_i$ for all $i$ under our setup. Therefore, given the constraints of (77), the unique optimal exists and is achieved. To observe the lack of duality gap, note that there exists a $\boldsymbol{\mu} \in \mathcal{M}$ and a $\boldsymbol{\lambda} \in [0,1]^n$ that is strictly feasible. Therefore, Slater's condition will imply lack of duality gap. We defer the proof of uniqueness of the dual optimal solution till a little later.

*Proof of (1).* Given the dual feasible $\mathbf{r} \in \mathbb{R}_+^n$, the structure of let $\boldsymbol{\mu}(\mathbf{r}), \boldsymbol{\lambda}(\mathbf{r})$ be the corresponding primal feasible solutions that maximize the Lagrangian, $\mathcal{L}$. Given structure of $\mathcal{L}$ as in (78), it follows that $\boldsymbol{\lambda}(\mathbf{r})$ must be such that

$$\lambda_i(\mathbf{r}) = \arg\max_{y \in [0,1]} \left( \beta U_i(y) - r_i y \right), \quad \text{for all } i.$$

For $\boldsymbol{\mu}(\mathbf{r})$, observe that

$$\frac{\partial \mathcal{L}(\boldsymbol{\mu}, \boldsymbol{\lambda}; \mathbf{r})}{\partial \mu_{\boldsymbol{\sigma}}} = -\log \mu_{\boldsymbol{\sigma}} - 1 + \boldsymbol{\sigma} \cdot \mathbf{r}.$$

Since $\boldsymbol{\mu}(\mathbf{r})$ is maximizing $\mathcal{L}$, from above it follows that $\mu_{\boldsymbol{\sigma}}(\mathbf{r}) \in (0,1)$ for all $\boldsymbol{\sigma} \in \mathcal{I}(G)$. Therefore, for any $\boldsymbol{\sigma}, \boldsymbol{\rho} \in \mathcal{I}(G)$ and $\boldsymbol{\sigma} \neq \boldsymbol{\rho}$, it must be that

$$\frac{\partial \mathcal{L}(\boldsymbol{\mu}(\mathbf{r}), \boldsymbol{\lambda}(\mathbf{r}); \mathbf{r})}{\partial \mu_{\boldsymbol{\sigma}}} = \frac{\partial \mathcal{L}(\boldsymbol{\mu}(\mathbf{r}), \boldsymbol{\lambda}(\mathbf{r}); \mathbf{r})}{\partial \mu_{\boldsymbol{\rho}}}.$$

That is,

$$\mu_{\boldsymbol{\sigma}}(\mathbf{r}) \propto \exp(\boldsymbol{\sigma} \cdot \mathbf{r}), \quad \text{for all} \quad \boldsymbol{\sigma} \in \mathcal{I}(G).$$

Thus, $\boldsymbol{\mu}(\mathbf{r}) = \boldsymbol{\pi}^{\mathbf{r}}$.

*Proof of (2).* Given (1), it follows that

$$\mathcal{D}(\mathbf{r}) = \mathcal{L}(\boldsymbol{\mu}(\mathbf{r}), \boldsymbol{\lambda}(\mathbf{r}); \mathbf{r}).$$

Now the dual variables $\mathbf{r}$ capture 'slack' in the corresponding constraints of (77). Specifically, for a given $\mathbf{r}$ if the corresponding primal solutions are $\boldsymbol{\mu}(\mathbf{r}), \boldsymbol{\lambda}(\mathbf{r})$, then the slack in the $i$th constraint is $s_i(\mathbf{r}) - \lambda_i(\mathbf{r})$: if it is positive, $r_i$ should be decreased and if it is negative, $r_1$ should be increased. This intuition is formalized in the optimization theory (e.g. see book by Boyd and Vandenberghe [7]) by establishing that a subgradient of the dual optimization at $\mathbf{r}$ is given by vector $g(\mathbf{r}) \in \mathbb{R}^n$ with

$$g_i(\mathbf{r}) = s_i(\mathbf{r}) - \lambda_i(\mathbf{r}).$$

*Proof of (3).* The uniqueness of solution of (77) was already explained. To understand uniqueness of $\mathbf{r}^*$, consider independent set $e_i$, which has only node $i$ in it; and the null set $\mathbf{0}$. Then, since $\boldsymbol{\mu}(\mathbf{r}^*) = \boldsymbol{\pi}^{\mathbf{r}^*}$ it follows that

$$\mu_{e_i}(\mathbf{r}^*) = \mu_{e_0}(\mathbf{r}^*)\exp(r_i^*).$$

Now suppose to contrary that there is another optimal solution of (80), $\widehat{\mathbf{r}} \neq \mathbf{r}^*$. Then, it will immediately contradict above as $\boldsymbol{\mu}^*$ is unique as discussed above. This completes the proof of (3).

**Convergence of $\mathbf{r}(j), \boldsymbol{\lambda}(j)$.** In light of Lemma 18(2), it follows that the algorithm (8) is motivated by the standard projected dual subgradient algorithm. The algorithm uses estimated $\widehat{\mathbf{s}}(\mathbf{r}(j))$ in place of $\mathbf{s}(\mathbf{r}(j))$; but exact update for $\boldsymbol{\lambda}(\mathbf{r}(j))$. That is, for all $i$

$$r_i(j+1) = [r_i(j) + \alpha(j)(\lambda_i(j) - \widehat{s}_i(j))]_+.$$

To this end, define 'error' vector

$$e(j) \stackrel{\triangle}{=} -\widehat{\mathbf{s}}(j) + \mathbf{s}(\mathbf{r}(j)).$$

And, let

$$d(j) \stackrel{\triangle}{=} \|\mathbf{r}(j) - \mathbf{r}^*\|_2^2.$$

Now consider the relation between $d(j+1)$ and $d(j)$. Since the projection $[\cdot]_+$ is non-expansive,

$$\begin{aligned} d(j+1) &\leq \|\mathbf{r}(j) - \mathbf{r}^* + \frac{1}{j}[\boldsymbol{\lambda}(j) - \mathbf{s}(\mathbf{r}(j)) + e(j)]\|_2^2 \\ &\leq d(j) + 2[\mathbf{r}(j) - \mathbf{r}^*]^T \cdot \frac{1}{j}[\boldsymbol{\lambda}(j) - \mathbf{s}(\mathbf{r}(j)) + e(j)] + O\left(\frac{n}{j^2}\right), \end{aligned}$$

where we have used the fact that each component of $\boldsymbol{\lambda}(j) - \mathbf{s}(\mathbf{r}(j)) + e(j)$ is $O(1)$. Define, the error in optimal cost at the $j$th step as

$$\Delta(j) = \mathcal{D}(\mathbf{r}(j)) - \mathcal{D}(\mathbf{r}^*).$$

By definition, $\Delta(j) \geq 0$. Since the dual objective $\mathcal{D}$ is convex, and $\mathbf{s}(\mathbf{r}(j)) - \boldsymbol{\lambda}(j)$ is its subgradient at $\mathbf{r}(j)$, we have

$$[\mathbf{r}(j) - \mathbf{r}^*]^T \cdot [\boldsymbol{\lambda}(j) - \mathbf{s}(\mathbf{r}(j))] \leq -\Delta(j). \tag{83}$$

Also, as used earlier, $r_i(j) = O(\log j)$ for all $i$. Therefore, from above we obtain that

$$d(j+1) \leq d(j) - \frac{2\Delta(j)}{j} + O\left(\frac{\log(j) + |\mathbf{r}^*|_\infty}{j}\|e(j)\|_1\right) + O\left(\frac{n}{j^2}\right). \tag{84}$$

Note that the analysis of Lemma 10 applies to bound $\|e(j)\|_1$ as is. That is,

$$\mathbb{E}[\|e(j)\|_1] = O\left(\frac{n}{j^2}\right). \tag{85}$$

Using this and taking expectation on both sides of inequality (84), we obtain

$$\mathbb{E}\left[d(j+1)\right] \leq \mathbb{E}\left[d(j)\right] - \frac{2}{j}\mathbb{E}\left[\Delta(j)\right] + O\left(\frac{n \cdot (\log(j) + |\mathbf{r}^*|_\infty)}{j^3}\right) + O\left(\frac{n}{j^2}\right).$$

Summing the above inequality from 1 to $\infty$, it follows that

$$\begin{aligned} 0 &\leq \mathbb{E}\left[d(\infty)\right] \\ &\leq \mathbb{E}\left[d(1)\right] - \left(\sum_{j=1}^{\infty}\frac{2}{j}\mathbb{E}\left[\Delta(j)\right]\right) + O(n). \end{aligned}$$

By rearranging the terms and using $E\left[d(1)\right] < \infty$, it follows that $\sum_{j=1}^{\infty} \frac{1}{j}\mathbb{E}\left[\Delta(j)\right] < \infty$. Since $\sum_{j=1}^{\infty} \frac{1}{j} = \infty$, we can conclude that

$$\begin{aligned} \liminf_{j} \mathbb{E}[\Delta(j)] = 0 &\Rightarrow \liminf_{j} \Delta(j) = 0, \quad \text{with probability 1} \\ &\Rightarrow \liminf_{j} \|\mathbf{r}(j) - \mathbf{r}^*\| = 0, \quad \text{with probability 1,} \end{aligned} \quad (86)$$

where we have used the fact that dual optimization (80) has a unique solution and it is convex minimization problem. Now, rest of the proof of $\mathbf{r}(j) \to \mathbf{r}^*(j)$ with probability 1 follows exactly the same set of arguments as those used in the proof of Theorem 1. The convergence of $\boldsymbol{\lambda}(j) \to \boldsymbol{\lambda}(\mathbf{r}^*) = \bar{\boldsymbol{\lambda}}$ follows due to continuity of solution of concave maximization (82) with respect to $\mathbf{r}$.

**Utility of $\bar{\boldsymbol{\lambda}}$, rate stability.** To begin with, we observe that convergence $\mathbf{r}(j) \to \mathbf{r}^*(j)$ and $\boldsymbol{\lambda}(j) \to \boldsymbol{\lambda}(\mathbf{r}^*) = \bar{\boldsymbol{\lambda}}$ with probability 1 implies the rate stability using exactly the same arguments as those used in Lemma 12.

To establish goodness of the $\bar{\boldsymbol{\lambda}}$, note that it along with $\boldsymbol{\mu}^*$ optimizes (77). Now $\boldsymbol{\lambda}^*$, the optimal allocation (as per (3)) along with appropriate distribution, say $\boldsymbol{\nu}^*$ on $\mathcal{I}(G)$) is a feasible solution. Therefore, it follows that

$$\begin{aligned} \beta \sum_{i} U_i(\lambda_i^*) &\leq \mathrm{H}_{ER}(\boldsymbol{\nu}^*) + \beta \sum_{i} U_i(\lambda_i^*) \\ &\leq \mathrm{H}_{ER}(\boldsymbol{\mu}^*) + \beta \sum_{i} U_i(\bar{\lambda}_i) \\ &\leq \log |\mathcal{I}(G)| + \beta \sum_{i} U_i(\bar{\lambda}_i). \end{aligned} \quad (87)$$

In above, we have used the fact that the entropy is non-negative and the maximum value of a discrete valued random variable's entropy is at most the logarithm of the cardinality of the support set. The (87) immediately implies the desired result. This completes the proof of Theorem 3.

## 6.2 Proof of Theorem 4

The proof of Theorem 4 in a nutshell requires us to establish that the average rate allocation $\tilde{\boldsymbol{\lambda}}$ has near optimal total utility. This follows using similar arguments that we used in proving

Theorem 3. That is, establish that the $\tilde{\boldsymbol{\lambda}}$ ends up approximately solving optimization problem (77). This property follows primarily because the congestion control algorithm 2 with update (11) is primarily designed as a constant step-size dual 'subgradient' algorithm. We will formalize this in the rest of this section. We begin with a useful property that establishes uniform bound on components of $\mathbf{r}(\cdot)$ and subsequently implies uniform bound on the components of the queue-size vector $\mathbf{Q}(\cdot)$ for all time duration. This will be followed by proof of the goodness of average rate $\tilde{\boldsymbol{\lambda}}$ to conclude the proof of Theorem 4.

**Uniform bound on $\|\mathbf{r}(j)\|_\infty$.** We state and prove the following bound on $\|\mathbf{r}(j)\|_\infty$ starting with $\mathbf{r}(0) = \mathbf{0}$.

**Lemma 19** *Under the update rule* (11), *for all* $1 \leq i \leq n$

$$r_i(j) \in [0, \beta V + \alpha], \quad \text{for all } j,$$

*where recall that $V$ is defined in* (10) *and $\alpha$ is the constant step-size used in the update* (11).

*Proof.* To prove this Lemma, consider any $i, 1 \leq i \leq n$. Now for any $j$, $r_i(j) \geq 0$ by the definition (cf. (11)). To prove $r_i(j) \leq \beta V + \alpha$, we will use the principle of mathematical induction. To this end, for the base case, $j = 0$ and $r_i(0) = 0$ by definition. Suppose, as the inductive hypothesis the property $r_i(j) \leq \beta V + \alpha$ is true for all $j \leq J$. Now we wish to establish this property for $j = J + 1$. To this end, we consider two cases: (a) $r_i(J) \leq \beta V$, or (b) $r_i(J) \in (\beta V, \beta V + \alpha]$.

First consider case (a). By (11), it follows that

$$\begin{aligned} r_i(J+1) &= [r_i(J) - \alpha \widehat{s}_i(J)]_+ + \alpha \lambda_i(J) \\ &\leq r_i(J) + \alpha \lambda_i(J) \\ &\leq r_i(J) + \alpha \\ &\leq \beta V + \alpha. \end{aligned}$$

In above we have used the fact that $\lambda_i(J) \in [0, 1]$ by definition. Now consider case (b). For this note that if $r_i(J) \in [\beta V, \beta V + \alpha]$, then the $\lambda_i(J) = 0$. This is because by (11), $\lambda_i(J)$ solves

$$\lambda_i(J) \in \arg\max_{y \in [0,1]} \{\beta U_i(y) - r_i(J) y\}, \tag{88}$$

and for any $y \in [0, 1]$,

$$\begin{aligned} \frac{d}{dy}(\beta U_i(y) - r_i(J) y) &= \beta U_i'(y) - r_i(J) \\ &\leq \beta V - r_i(J) \\ &< 0. \end{aligned} \tag{89}$$

That is, the optimal solution of (88) is 0. This completes the proof of Lemma 19. □

**Uniform bound on $\|\mathbf{Q}(j)\|_\infty$.** We state and prove the following bound on $\|\mathbf{Q}(j)\|_\infty$ starting with $\mathbf{Q}(0) = \mathbf{0}$.

**Lemma 20** *Under the congestion control algorithm 2, starting with empty queue, i.e.* $\mathbf{Q}(0) = \mathbf{0}$, *the following hold for all* $t \geq 0$:

$$Q_i(t) \leq \frac{T}{\alpha}\left(\beta V + 2\alpha\right).$$

*Proof.* In what follows, we will show that for time instances $t = jT$, for $j \geq 0$, the queue-size is bounded as

$$Q_i(jT) \leq \frac{T}{\alpha} r_i(j), \quad \text{for all } i. \tag{90}$$

The (90) along with the bound on $r_i(\cdot)$ implied by Lemma 19, will imply

$$Q_i(jT) \leq \frac{T}{\alpha}\left(\beta V + \alpha\right), \quad \text{for all } i. \tag{91}$$

Finally, by noticing that $\lambda_i(j) \in [0,1]$ for all $i,j$, it follows that for any $t \in [jT, (j+1)T)$, $Q_i(t) \leq Q_i(jT) + T$. Therefore, we will obtain the desired result of Lemma 20.

Now we prove the remaining bounded as stated in (90). To this end, note that

$$Q_i((j+1)T) \leq [Q_i(jT) - \widehat{s}_i(j)T]_+ + \lambda_i(j)T. \tag{92}$$

This follows by imagining that all the arrival traffic in $[jT, (j+1)T)$, $\lambda_i(j)T$ amount of data, is added to the queue at the end of the interval; service $\widehat{s}_i(j)T$ is used only to serve data that was present at time $jT$.

Based on (92), we will establish (90), by means of the principle of mathematical induction. For the based case of $j = 0$, we have $Q_i(0) = 0$ and $r_i(0) = 0$. For induction hypothesis, assume it to hold true for all $j \leq J$. For $j = J + 1$, we wish to establish that the relation holds. To this end, using (92) it follows that

$$\begin{aligned} Q_i((J+1)T) &\leq [Q_i(JT) - \widehat{s}_i(J)T]_+ + \lambda_i(J)T \\ &\leq \left[\frac{T}{\alpha} r_i(J) - \widehat{s}_i(J)T\right]_+ + \lambda_i(J)T \\ &= \frac{T}{\alpha}\left([r_i(J) - \alpha \widehat{s}_i(J)]_+ + \alpha \lambda_i(J)\right) \\ &= \frac{T}{\alpha} r_i(J+1). \end{aligned} \tag{93}$$

Here the last equality follows by definition (11). This completes the proof of (90) and Lemma 20. □

**A useful variational characterization.** We state the Gibbsian variational characterization (e.g. see book [15]) of the distribution $\boldsymbol{\pi}^{\mathbf{r}}$ that will be useful later in the proof.

**Lemma 21** *Given* $\mathbf{r} \in \mathbb{R}^n$, $\boldsymbol{\pi}^{\mathbf{r}}$ *is the unique solution of*

$$\begin{aligned} \text{maximize} \quad & \mathbb{E}_{\boldsymbol{\mu}}[\boldsymbol{\sigma} \cdot \mathbf{r}] + H_{ER}(\boldsymbol{\mu}) \\ \text{over} \quad & \boldsymbol{\mu} \in \mathcal{M}, \end{aligned} \tag{94}$$

where recall that $\mathcal{M}$ is the space of probability distributions over $\mathcal{I}(G)$. Further,

$$\mathbb{E}_{\pi^{\mathbf{r}}}[\boldsymbol{\sigma}\cdot\mathbf{r}] \geq \max_{\boldsymbol{\lambda}\in\Lambda}\boldsymbol{\lambda}\cdot\mathbf{r} - \log|\mathcal{I}(G)|. \qquad (95)$$

*Proof.* The (94) was established implicitly in Lemma 18. To see an explicit proof, consider the following. For any $\boldsymbol{\mu}\in\mathcal{M}$,

$$\begin{aligned}
\mathbb{E}_{\boldsymbol{\mu}}[\boldsymbol{\sigma}\cdot\mathbf{r}] + \mathrm{H}_{ER}(\boldsymbol{\mu}) &= \sum_{\boldsymbol{\sigma}\in\mathcal{I}(G)} (\boldsymbol{\sigma}\cdot\mathbf{r})\mu_{\boldsymbol{\sigma}} - \sum_{\boldsymbol{\sigma}\in\mathcal{I}(G)} \mu_{\boldsymbol{\sigma}}\log\mu_{\boldsymbol{\sigma}} \\
&\stackrel{(a)}{=} \sum_{\boldsymbol{\sigma}\in\mathcal{I}(G)} (\log \pi^{\mathbf{r}}_{\boldsymbol{\sigma}} + \log Z(\mathbf{r}))\mu_{\boldsymbol{\sigma}} - \sum_{\boldsymbol{\sigma}\in\mathcal{I}(G)} \mu_{\boldsymbol{\sigma}}\log\mu_{\boldsymbol{\sigma}} \\
&= \log Z(\mathbf{r}) + \sum_{\boldsymbol{\sigma}\in\mathcal{I}(G)} \mu_{\boldsymbol{\sigma}}\log\frac{\pi^{\mathbf{r}}_{\boldsymbol{\sigma}}}{\mu_{\boldsymbol{\sigma}}} \\
&\stackrel{(b)}{\leq} \log Z(\mathbf{r}). \qquad (96)
\end{aligned}$$

In above (a) follows from the fact that

$$\pi^{\mathbf{r}}_{\boldsymbol{\sigma}} = \frac{1}{Z(\mathbf{r})}\exp\left(\boldsymbol{\sigma}\cdot\mathbf{r}\right).$$

The (b) follows from an application of Jensen's inequality. The above suggests that, the optimal cost of (94) is $\log Z(\mathbf{r})$ and is achieved iff the $\boldsymbol{\mu} = \boldsymbol{\pi}^{\mathbf{r}}$. This establishes the first claim of Lemma 21.

To see (95), define $\boldsymbol{\mu}^*$ as

$$\mu^*_{\boldsymbol{\sigma}} = \begin{cases} 1 & \text{if } \boldsymbol{\sigma} = \boldsymbol{\sigma}^* \\ 0 & \text{o.w.} \end{cases}$$

Here $\boldsymbol{\sigma}^* = \arg\max_{\boldsymbol{\sigma}\in\mathcal{I}(G)} \boldsymbol{\sigma}\cdot\mathbf{r}$. Then, using the above it follows that

$$\begin{aligned}
\mathbf{s}(\mathbf{r})\cdot\mathbf{r} &= \mathbb{E}_{\pi^{\mathbf{r}}}[\boldsymbol{\sigma}\cdot\mathbf{r}] \\
&\geq \mathbb{E}_{\boldsymbol{\mu}^*}[\boldsymbol{\sigma}\cdot\mathbf{r}] + \mathrm{H}_{ER}(\boldsymbol{\mu}^*) - \mathrm{H}_{ER}(\boldsymbol{\pi}^{\mathbf{r}}) \\
&\stackrel{(a)}{\geq} \boldsymbol{\sigma}^*\cdot\mathbf{r} + 0 - \log|\mathcal{I}(G)| \\
&\stackrel{(b)}{=} \max_{\boldsymbol{\lambda}\in\Lambda}\boldsymbol{\lambda}\cdot\mathbf{r} - \log|\mathcal{I}(G)|. \qquad (97)
\end{aligned}$$

In above (a) follows from the definition of $\boldsymbol{\mu}^*$ and the fact that for any distribution on $\mathcal{M}$, the entropy is at the most $\log|\mathcal{I}(G)|$. The (b) follows because any $\boldsymbol{\lambda}\in\Lambda$ is a convex combination of elements in $\mathcal{I}(G)$. $\square$

**Some properties.** Here we state some useful properties that will be useful in completing proof of Theorem 4. To begin with, let $\boldsymbol{\lambda}^*$ be the optimal solution to congestion control problem (3). At any stage $j$, $\boldsymbol{\lambda}(j)$ is obtained as

$$\lambda_i(j) \in \arg\max_{y\in[0,1]} \{\beta U_i(y) - r_i(j)y\}, \qquad \text{for all } i.$$

Therefore, it follows that for any $j$

$$\beta U_i(\lambda_i(j)) - r_i(j)\lambda_i(j) \geq \beta U_i(\lambda_i^*) - r_i(j)\lambda_i^*. \tag{98}$$

Since $\boldsymbol{\lambda}^* \in \boldsymbol{\Lambda}$, we have

$$\sum_i \lambda_i^* r_i(j) \leq \max_{\boldsymbol{\lambda} \in \boldsymbol{\Lambda}} \boldsymbol{\lambda} \cdot \mathbf{r}(j). \tag{99}$$

Define notation $m^*(\mathbf{r}) = \max_{\boldsymbol{\lambda} \in \boldsymbol{\Lambda}} \boldsymbol{\lambda} \cdot \mathbf{r}$. From (98) and (99), we have

$$\mathbf{r}(j) \cdot \boldsymbol{\lambda}(j) \leq \beta\left(\sum_i U_i(\lambda_i(j))\right) - \beta\left(\sum_i U_i(\lambda_i^*)\right) + m^*(\mathbf{r}(j)). \tag{100}$$

We will observe another useful property. By Lemma 19, we have $\|\mathbf{r}(j)\|_\infty$ bounded by $\beta V + \alpha$. Therefore, using the mixing time bounds and arguments utilized in Lemma 10, we obtain that by the choice of appropriately large $T$ as

$$T = \exp\left(\Theta(\beta n V)\right) \Theta\left(\frac{(\beta V + \alpha)n^2}{\beta \varepsilon}\right), \tag{101}$$

we have that for all $j$,

$$|\mathbb{E}[\widehat{s}_i(j)|\mathcal{F}_j] - s_i(\mathbf{r}(j))| \leq \frac{\beta \varepsilon}{10(\beta V + \alpha)n}, \quad \text{for all } i. \tag{102}$$

In above, the conditioning $\mathcal{F}_j$ represents the filteration (or information) till time $L(j)$; while recall that the random variable $\widehat{s}_i(j)$ is the empirical service rate in $[L(j), L(j+1))$.

**Wrapping up: Completing proof of Theorem 4.** Now, let us start with the algorithm's update rule (11). Specifically, for a given $i$, squaring both sides of (11) for $r_i(\cdot)$ gives us

$$\begin{aligned}
r_i^2(j+1) &= \left([r_i(j) - \alpha \widehat{s}_i(j)]_+ + \alpha \lambda_i(j)\right)^2 \\
&= [r_i(j) - \alpha \widehat{s}_i(j)]_+^2 + 2\alpha \lambda_i(j)[r_i(j) - \alpha \widehat{s}_i(j)]_+ + \alpha^2 \lambda_i^2(j) \\
&\stackrel{(a)}{\leq} [r_i(j) - \alpha \widehat{s}_i(j)]^2 + 2\alpha \lambda_i(j) r_i(j) + \alpha^2 \\
&\stackrel{(b)}{\leq} r_i(j)^2 + 2\alpha r_i(j)[\lambda_i(j) - \widehat{s}_i(j)] + 2\alpha^2.
\end{aligned} \tag{103}$$

In above (a) follows from the fact that $[x]_+^2 \leq x^2$ and $\lambda_i(j) \in [0,1]$ for all $i,j$; and (b) follows from the fact that $\widehat{s}_i(j) \in [0,1]$ for all $i,j$. From (103), we have that

$$\begin{aligned}
\left(\sum_i r_i^2(j+1) - r_i^2(j)\right) &\leq 2\alpha \left(\sum_i r_i(j)(\lambda_i(j) - \widehat{s}_i(j))\right) + 2n\alpha^2 \\
&= 2\alpha \left[\sum_i r_i(j)(\lambda_i(j) - s_i(\mathbf{r}(j))) + \sum_i r_i(j)(s_i(\mathbf{r}(j)) - \widehat{s}_i(j)) + n\alpha\right] \\
&= 2\alpha \left[\mathbf{r}(j) \cdot \boldsymbol{\lambda}(j) - \mathbf{r}(j) \cdot \mathbf{s}(\mathbf{r}(j)) + \mathbf{r}(j) \cdot (\mathbf{s}(\mathbf{r}(j)) - \widehat{\mathbf{s}}(j)) + n\alpha\right]. \tag{104}
\end{aligned}$$

By (97) and since $|\mathcal{I}(G)| \leq 2^n$, we have

$$-2\alpha\, \mathbf{r}(j) \cdot \mathbf{s}(\mathbf{r}(j)) \;\leq\; -2\alpha\, m^*(\mathbf{r}(j)) + 2\alpha\, n. \tag{105}$$

Therefore, using (100) we have

$$2\alpha\, \mathbf{r}(j) \cdot \boldsymbol{\lambda}(j) - 2\alpha\, \mathbf{r}(j) \cdot \mathbf{s}(\mathbf{r}(j)) \;\leq\; 2\alpha\beta\left(\sum_i U_i(\lambda_i(j)) - U_i(\lambda_i^*)\right) + 2\alpha n. \tag{106}$$

Now using (106) in (104) and fact that $\alpha^2 \leq \alpha$ because $\alpha \in (0,1)$, we have

$$\left(\sum_i r_i^2(j+1) - r_i^2(j)\right) \;\leq\; 2\alpha\beta(U(\boldsymbol{\lambda}(j)) - U(\boldsymbol{\lambda}^*)) + 2\alpha\, \mathbf{r}(j) \cdot (\mathbf{s}(\mathbf{r}(j)) - \widehat{\mathbf{s}}(j)) + 4\alpha n, \tag{107}$$

where we have used notation $U(\boldsymbol{\lambda}) = \sum_i U_i(\lambda_i)$. Now taking its summation from $j = 0$ till $J-1$ on both sides of (107), the fact that $\mathbf{r}(0) = \mathbf{0}$ and diving both side by $J$, we have

$$\frac{1}{J}\sum_i r_i^2(J) \;\leq\; 2\alpha\beta\left(\sum_{j=0}^{J-1}\frac{U(\boldsymbol{\lambda}(j))}{J} - U(\boldsymbol{\lambda}^*)\right) + \frac{1}{J}\left(\sum_{j=0}^{J-1} 2\alpha\mathbf{r}(j) \cdot (\mathbf{s}(\mathbf{r}(j)) - \widehat{\mathbf{s}}(j))\right) + 4\alpha n. \tag{108}$$

Now, define $\Delta(j) = 2\alpha\mathbf{r}(j) \cdot (\mathbf{s}(\mathbf{r}(j)) - \widehat{\mathbf{s}}(j))$ and $X(j) = \Delta(j) - \mathbb{E}[\Delta(j)|\mathcal{F}_j]$. By definition, $S(j) = \sum_{k=0}^{j-1} X(k)$ is a martingale with respect to filteration $\{\mathcal{F}_j\}_{j \geq 1}$. With this notation, we have that for any $J$,

$$\frac{1}{J}\left(\sum_{j=0}^{J-1} 2\alpha\mathbf{r}(j) \cdot (\mathbf{s}(\mathbf{r}(j)) - \widehat{\mathbf{s}}(j))\right) \;=\; \frac{1}{J}\left(\sum_{j=0}^{J-1} X(j) + \mathbb{E}[\Delta(j)|\mathcal{F}_j]\right)$$
$$\stackrel{(a)}{\leq} \frac{1}{J}S(J) + \frac{\alpha\beta\varepsilon}{5}. \tag{109}$$

In above (a) follows from (102) and bound on $\mathbf{r}(\cdot)$ using Lemma 19. Finally, note that $S(\cdot)$ is a martingale with bounded increment due to uniform bound on $\mathbf{r}(\cdot)$, the fact that $\mathbf{s}(\cdot), \widehat{\mathbf{s}}(\cdot)$ are vectors in $[0,1]^n$ and $\alpha \in (0,1)$. Therefore, by Strong Law of Large Large Numbers for martingales with bounded increments it follows that

$$\lim_{J \to \infty} \frac{1}{J} S(J) \;=\; 0, \qquad \text{with probability 1.} \tag{110}$$

That is, with probability 1

$$\limsup_{J \to \infty} \frac{1}{J}\left(\sum_{j=0}^{J-1} 2\alpha\mathbf{r}(j) \cdot (\mathbf{s}(\mathbf{r}(j)) - \widehat{\mathbf{s}}(j))\right) \;\leq\; \frac{\alpha\beta\varepsilon}{5}. \tag{111}$$

Using (111) in (108) along with Lemma 19, and then taking $J \to \infty$, we have that with probability 1,

$$\liminf_{J \to \infty} \sum_{j=0}^{J-1} \frac{U(\boldsymbol{\lambda}(j))}{J} \;\geq\; U(\boldsymbol{\lambda}^*) - \frac{\varepsilon}{10} - \frac{2n}{\beta}. \tag{112}$$

Finally, observe that by concavity of function $\sum_i U_i(\cdot)$ along with Jensen's inequality, we have that for $\tilde{\lambda}_i(J) = (\sum_{j=0}^{J-1} \lambda_i(j))/J$,

$$U(\tilde{\boldsymbol{\lambda}}(J)) \geq \sum_{j=0}^{J-1} \frac{U(\boldsymbol{\lambda}(j))}{J}.$$

Therefore, the following desired conclusion of Theorem 4 follows from (112) along with choice of $\beta = 4n/\varepsilon$: with probability 1,

$$\liminf_{J \to \infty} U(\tilde{\boldsymbol{\lambda}}(J)) \geq U(\boldsymbol{\lambda}^*) - \varepsilon. \tag{113}$$

# 7 Conclusion

In this paper, we have presented a simple, distributed randomized algorithm for scheduling and congestion control in a network. Our algorithm is essentially a random access protocol with time-varying access probabilities. Our algorithm for scheduling, in the presence of exogeneous arrivals, achieves throughput optimality while our algorithm for scheduling with congestion controlled arrivals achieves near-optimal resource allocation when nodes have concave utilities. We believe that the algorithmic method presented in this paper should be of general interest.